\documentclass[nofootinbib, twocolumn, prb, aps, 10pt]{revtex4-2}
\usepackage[lmargin=1.55cm, rmargin=1.45cm, vmargin=1.75cm]{geometry}
\usepackage[utf8]{inputenc}
\DeclareUnicodeCharacter{202F}{\,}
\usepackage{longtable}
\usepackage{xcolor}
\usepackage{blindtext,alltt}
\usepackage{amsmath}
\usepackage{amssymb}
\usepackage{mathrsfs}
\usepackage{amsfonts}
\usepackage{amssymb}
\usepackage{graphicx}
\usepackage{mathrsfs}
\usepackage{hyperref}
\usepackage{mwe}
\usepackage{bbold}
\usepackage{soul}
\graphicspath{{Img/}}

\newcommand{\e}{\textrm{e}}
\newcommand{\bra}[1]{\big \langle #1 \big|}
\newcommand{\ket}[1]{\big | #1 \big \rangle}

\newcommand{\melem}[3]{\big \langle #1 \big | #2 \big| #3 \big \rangle}
\newcommand{\Ham}{\widehat{\mathrm{H}}}
\newcommand{\MC}{\, \overset{\textrm{M.C.}}{=}\, }

\newcommand{\U}{\mathcal{U}}
\newcommand{\id}{\mathbb{1}}
\newcommand{\I}{\mathrm{I}}
\newcommand{\PP}{\mathrm{P}}
\newcommand{\pp}{\mathrm{p}}

\newcommand{\llangle}{\langle\!\langle}
\newcommand{\rrangle}{\rangle\!\rangle}

\begin{document}
	
	\title{Multi-particle quantum systems within the Worldline Monte Carlo formalism}

	\author{Ivan Ahumada}
	\email{ivan.ahumadahernandez@plymouth.ac.uk}
	\affiliation{Centre for Mathematical Sciences, University of Plymouth, Plymouth, PL4 8AA, UK}

	\author{Filippo Ricchetti}
	\email{filippo.ricchetti.physics@gmail.com}
	\affiliation{Dipartimento di Scienze Fisiche, Informatiche e Matematiche, Universit\`a degli Studi di Modena e Reggio Emilia, Via Campi 213/a, 41125 Modena, Italy}

	\author{Federico Grasselli}
	\email{federico.grasselli@unimore.it}
	\affiliation{Dipartimento di Scienze Fisiche, Informatiche e Matematiche, Universit\`a degli Studi di Modena e Reggio Emilia, Via Campi 213/a, 41125 Modena, Italy}

	\author{Olindo Corradini}
	\email{olindo.corradini@unimore.it}
	\affiliation{Dipartimento di Scienze Fisiche, Informatiche e Matematiche, Universit\`a degli Studi di Modena e Reggio Emilia, Via Campi 213/a, 41125 Modena, Italy}
    \affiliation{I.N.F.N., Sezione di Bologna, Via Irnerio 46, 40126 Bologna, Italy}

	\author{Guido Goldoni}
	\email{guido.goldoni@unimore.it}
	\affiliation{Dipartimento di Scienze Fisiche, Informatiche e Matematiche, Universit\`a degli Studi di Modena e Reggio Emilia, Via Campi 213/a, 41125 Modena, Italy}
	
	\author{Marco A. Palomino}
	\email{marco.palomino@abdn.ac.uk}
	\affiliation{School of Natural and Computing Sciences, University of Aberdeen, Aberdeen, AB24 3UE, UK}

	\author{Max Badcott}
	\email{max.badcott@students.plymouth.ac.uk}
	\affiliation{Centre for Mathematical Sciences, University of Plymouth, Plymouth, PL4 8AA, UK}
	
	\author{James P. Edwards}
	\email{james.p.edwards@plymouth.ac.uk}
	\affiliation{Centre for Mathematical Sciences, University of Plymouth, Plymouth, PL4 8AA, UK}

        	\author{Craig McNeile}
	\email{craig.mcneile@plymouth.ac.uk}
	\affiliation{Centre for Mathematical Sciences, University of Plymouth, Plymouth, PL4 8AA, UK}

	\begin{abstract}
		We extend the Worldline Monte Carlo approach to computationally simulating the Feynman path integral of non-relativistic multi-particle quantum-mechanical systems.
        %in non-relativistic quantum mechanics to multi-particle systems.
        We show how to generate an arbitrary number of worldlines distributed according to the (free) kinetic part of the multi-particle quantum dynamics and how to simulate interactions between worldlines in the ensemble. We test this formalism with  two- and three-particle quantum mechanical systems,  with both long range Coulomb-like interactions between the particles and external fields acting separately on the particles, in various spatial dimensionality. We extract accurate estimations of the ground state energy of these systems using the late-time behaviour of the propagator, validating our approach with numerically exact solutions obtained via straightforward diagonalisation of the Hamiltonian. Systematic benchmarking of the new approach, presented here for the first time, shows that the computational complexity of Wordline Monte Carlo scales more favourably with respect to standard numerical alternatives. The method, which is general, numerically exact, and computationally not intensive, can easily be generalised to relativistic systems.
	\end{abstract}
	
	\maketitle	
	
	%%%%%%%%%%%%%%%%%%%%%%%%%%%%%%%%%%%%%%%%%%%%%%%%%%%%%%%%%%%%%%%%%%%%%%%%%%%%%%%%%%%%%%%%%%%%%%
	\section{Introduction}
	\label{sec:Intro}
    It is generally challenging to determine the states of a closed quantum mechanical system exactly -- unless it is particularly simple or enjoys special symmetries (a good example being the class with ``shape-invariant'' quantum potentials \cite{cooper2001supersymmetry}). Even for relatively ordinary potentials, analytic studies of the quantum mechanical system can be challenging; this is the case, for example, with the Yukawa potential \cite{Yukawa:1935xg, Jost:1951zz, Rogers:1970axl, Li:2006as, Edwards:2017ndv, Napsuciale:2021qtw}, where the exact ground state energy, wavefunction and critical coupling (the value of the screening parameter at which bound states cease to exist) remain unknown in closed form. As such, a wide variety of approximate and numerical techniques have been developed to study non-trivial quantum systems. 
	
	Amongst these are schemes which focus on the Hamiltonian formulation of the theory, including perturbation theory and the variational principle. Numerical approaches, based on discretising the quantum Hamiltonian, are of course also popular. Generally, numerically exact solutions of the Schrödinger equation can be provided to arbitrary accuracy in any dimensionality by straightforward, although possibly computationally intensive, matrix representations, e.g, in position-space or momentum-space, or via basis expansion \cite{giannozzi2013numerical}. Interacting multi-particle systems are often {exactly diagonalised} -- in the numerical sense -- by configuration interaction \cite{Rontani2006a} or coupled cluster methods \cite{Bartlett2007}, once single particle states are available. These techniques often imply large in-memory matrix storage and diagonalisation and tend not to scale well with particle number.
		
	Of course, an alternative, equivalent formulation of quantum mechanics exists, developed by Feynman and Dirac, the so-called ``path integral'' approach, which is a Lagrangian formulation rather than a Hamiltonian one.
    %favours the Lagrangian over the Hamiltonian. 
    Its key object of interest is in the Feynman propagator of the system, defined as the position space matrix elements of the time evolution operator, $\widehat{\U}$, which we review here for a non-relativistic system with one scalar degree of freedom:
	\begin{equation}
		K(x', x; T) := \melem{x'}{\widehat{\U}(T)}{x} \,, \quad \ket{\Psi(T)} = \widehat{\U}(T)\ket{\Psi(0)}\,.
	\end{equation}
	The propagator (or kernel) satisfies its own version of the Schrödinger equation ($\hbar = 1$),
	\begin{equation}
		i \frac{\partial \widehat{\U}(T)}{\partial T} = \Ham \widehat{\U}(T)\,, \qquad \widehat{\U}(0) = 1.
	\end{equation}
	Since the time evolution operator has spectral decomposition in terms of eigenstates/eigen-energies of the Hamiltonian, $\Ham$ (we assume a static Hamiltonian for simplicity),
	\begin{align}
		\hspace{-2em} \widehat{\U}(T) &= \e^{-i \Ham T} \\ 
        \hspace{-2em} &=  \sum_{n} \e^{-i E_{n}T} \ket{\psi_{n}}\bra{\psi_{n}} + \int dE\, \e^{-i ET} \ket{\psi_{E}}\bra{\psi_{E}}\,,
		\label{eq:Uasymp}
	\end{align}
	with the sum (respectively integral) running over bound (scattering) states of the theory, knowing the Feynman propagator is equivalent to solving the quantum mechanical system. Taking $T \rightarrow \infty(1 - i\epsilon)$ projects asymptotically onto the ground state of the system. For the propagator,
	\begin{equation}
		\lim_{T \to \infty(1 - i\epsilon)} K(x', x; T) \sim \psi_{0}(x')\psi_{0}^{\star}(x)\, \e^{-iE_{0}T}\,, \label{eq:PropEspectralDecomp}
	\end{equation}
	up to exponentially suppressed corrections (assuming a ``mass gap'' $E_{1} - E_{0} > 0$).

	The propagator admits a path-integral representation consisting of a weighted sum over trajectories with endpoints fixed at $x$ and $x'$. For a separable Hamiltonian of the form, $\Ham = \frac{\hat{p}^{2}}{2m} + V(\hat{x})$,
	\begin{align}
		K(x', x; T) &= \int_{x(0) = x}^{x(T) = x'} \hspace{-1.5em}\mathscr{D}x(\tau)\, \e^{i S[x(\tau)]}\,, \\ S[x(\tau)] :\!&= \int_{0}^{T}d\tau\, \Big[\frac{m\dot{x}^{2}}{2} - V(x(\tau)) \Big]\,,
	\end{align}
	where $S[x]$ is the classical configuration space action associated to each trajectory. Then, for the purpose of this article, we can manipulate this into a form suitable for Monte Carlo estimation by multiplying and dividing by the free-particle kernel, $K_{0}$, and rotating to the imaginary time formalism via $\tau \rightarrow -i\tau$ and $T \rightarrow -iT \equiv T_{E}$,
	\begin{align}
		K(x', x; T_{E}) &= K_{0}(x', x; T_{E}) \Big\langle \e^{- \int_{0}^{T_{E}}d\tau \, V(x(\tau))}  \Big\rangle \label{eq:PropWMC} \\
		\big\langle \hat{\mathcal{O}}  \big\rangle :\!&=\,  \frac{\displaystyle \int_{x(0) = x}^{x(T_{E}) = x'} \hspace{-1.5em}\mathscr{D}x(\tau)\, \hat{\mathcal{O}}\,\e^{-\int_{0}^{T_{E}} d\tau \frac{m\dot{x}^{2}}{2}}}{ \displaystyle \int_{x(0) = x}^{x(T_{E}) = x'} \hspace{-1.5em}\mathscr{D}x(\tau)\, \e^{-\int_{0}^{T_{E}} d\tau \frac{m\dot{x}^{2}}{2}} }\,, \label{eq:ExpectVal}
	\end{align}
	where we will henceforth drop the $E$ on the ``Euclidean time'' for brevity and we have normalised such  that $\langle 1 \rangle = 1$. Above, the (Heisenberg) \emph{operator} $\hat{\mathcal{O}}$ is generically a function of coordinates, $\hat{x}(\tau_1),\,\hat{x}(\tau_2),\dots$, and momenta (derivatives thereof). 
    
    For a numerical estimation of the expectation value $\big\langle\cdots\big\rangle$ we discretise the integral over paths to a finite sum over a sample, size $N_{L}$, of trajectories, $x_i(\tau)$, which are generated according to the Gaussian distribution on their velocities\footnote{Note that we are making the choice, here, to generate Brownian motion trajectories according to the \textit{free}-particle action. This aims at universality of application to any quantum system. Alternative approaches which estimate expectation values taking the potential into account also exist as described in the main text.},
    \begin{align}
       \mathscr{P}\big( \{ x_{i} \}\big) &\sim \e^{- \int_{0}^{T} d\tau \frac{m\dot{x}_{i}^{2}}{2} }\,.
		\label{eq:Px} 
    \end{align}
    Thus, the Monte Carlo version of the above expectation value reads
	\begin{align}
		\Big\langle \e^{- \int_{0}^{T}d\tau \, V(x(\tau))}  \Big\rangle &\MC \frac{1}{N_{L}} \sum_{i = 1}^{N_{L}} \e^{- \int_{0}^{T}d\tau \, V(x_{i}(\tau))}\,.
	\end{align}
	To estimate the remaining parameter integral, a further (Riemann) discretisation of the trajectory with respect to its  {time parameter}, $\tau$, reduces the problem to one of producing $N_{P}$ points with Gaussian distribution on their finite difference squared (more details and algorithms for generating trajectories are provided in the Appendix \ref{sec:ApAlg}) and evaluating the Wilson line along each such discretised path.

	Numerical simulations of the path integral in this form were first considered in \cite{TjonThesis,Nieuwenhuis:1995ux, Nieuwenhuis:1996mc}. These studies focused on bound state formation based on a proper-time formalism of quantum field theory called the ``Feynman-Schwinger representation'' \cite{Savkli:1999rw, Savkli:2002fj}. There, stochastic trajectories were generated with a heat-bath algorithm that thermalises the trajectories towards minimising the (full) action. This approach has several drawbacks: it must discard samples to (a) arrive at representative trajectories and (b) reduce (but not remove) autocorrelation between samples, and involves a probabilistic rejection of candidate trajectories, wasting further processing time. Later the algorithms were adapted to the worldline formalism of quantum field theory (a first quantised approach based on formulating the theory in terms of path integrals over relativistic point particles \cite{Strass1, SchmidtRev, ChrisRev, UsRep}), with trajectories generated directly with distribution (\ref{eq:Px}) \cite{Gies:2001zp, Gies:2001tj,  Schmidt:2002mt, Langfeld:2002vy, Gies:2003cv, Gies:2005sb, Langfeld:2007wh}. This approach has the advantage of no correlation between sample trajectories and is completely rejection free. The former approach is useful for calculating expectation values of operators weighted by the full action, but is computationally costly unless the algorithms are specialised to the system under consideration (recent studies using the Metropolis-Hastings algorithm can be found in \cite{Westbroek, Westbroek2, Marik:2022fzd,Creutz:1980gp}). The latter approach, dubbed worldline Monte Carlo (WMC), aims at universality, treating the interaction part of the action separately from the free kinetic term.
	Furthermore, in Hybrid Monte Carlo simulations of lattice field theory it has been found~\cite{Davies:1989vh,Horsley:2023fhj,Fields:2025ydm}
	that speed ups can be obtained by modifying the kinetic term of the Hamiltonian
	using Fourier acceleration and information from free field theory.
	The treatment of the free kinetic term in WMC may be similarly advantageous.
	
	Yet more recently, the same method has been applied to quantum mechanical propagators (i.e.~to the non-relativistic limit) \cite{UsMonteCarlo} and the heat kernel \cite{Franchino-Vinas:2019udt, Corradini:2020tgk} in an effort to calculate the energies and wavefunctions of low-lying states, and to fermionic models \cite{dunne2009worldline}. Statistical properties of the values of the Wilson line in the space of Brownian motion trajectories were studied in \cite{UsPv, nHit, PvHz}, and will be important in the later development of the algorithms used to sample multi-particle systems here. In particular, a late-time undersampling problem was effectively overcome for a wide class of systems in \cite{UsPv} by the introduction of ``background potentials.'' 
	%As we will show below, such an approach is ideally suited for the problem of studying excitons in the present manuscript. \IA{The above phrase is not longer true, perhaps we should consider removing it.}

	With the exception of some very early work (in any case based on thermalisation), little has been done to extend the WMC technique to multi-particle systems. However, the developments in references~~\cite{UsMonteCarlo,UsPv} have laid the foundations for returning to this setting, opening a new route to studying inter-particle interactions for both free and constrained systems. Of particular relevance for this article, in Ref.~~\cite{UsMonteCarlo} the authors adapted and developed a ``smoothing algorithm'' for dealing with singular potentials inspired by \cite{TjonThesis} -- this is important if inter-particle interactions have small-distance divergences, as encountered below.
    
	Here, we aim to generalise the WMC formalism to multi-particle systems, demonstrating its validity and range of applicability. As a challenging proving ground, we shall determine the bound ground state energy of two and three particle systems with an attractive interaction having a long-range Coulomb tail. Of course, for translationally invariant interaction potentials the two-particle problem in free space reduces to a trivial one-particle problem, due to the separation of the relative and centre of mass degrees of freedom (DoFs), but it is a non trivial problem in the presence of external potentials acting separately on the particles, which remove translational invariance and interlace the DoFs. Similarly, three-particle systems cannot be separated even in the free case. 
    
    While numerically amenable, two-particle systems are also of practical interest, e.g.,  in condensed matter physics, where Coulomb-bound electron-hole pairs -- aka \textit{excitons} -- are the most important interband light-coupled excitation. In an excitonic complex, an electron is  excited to the conduction band by light absorption, which interacts with the hole left behind in the valence band. In the simplest description, which is named the envelope function approach, although embedded in a material, electrons and holes can often be modelled as free particles in vacuum with strongly renormalised (and unequal) effective masses, bound by a reciprocal Coulomb interaction \cite{bastard1982exciton} that is  significantly screened by the relatively large dielectric constant of the material; hence, the pair is delocalised over distance much larger than the underlying material lattice spacing. In semiconductor language, this is called a Wannier-Mott exciton. Notably, excitonic complexes are particularly important in one or two dimensions (1D or 2D) \cite{WOS:A1996UW49000044,WOS:A1997WX51800027, bastard1982exciton}, as we study in the present paper, since a large part of opto-electronic technology is based on semiconductor nano-structures (see, e.g., \cite{Huang2004}), where one or more degree of freedom is  \textit{frozen} by lateral quantum confinement \cite{Bastard1990}. There, the particle dynamics is effectively 1D or 2D,  the exciton binding energy is larger, and light coupling is stronger.
    
    Excitons may occur as free complexes (their centre-of-mass moves as a free particle in the crystal lattice) or be localised by an electrostatic potential generated by crystal defects, impurities, interfaces etc. In nano-devices, external electrostatic fields generated by properly engineered gates may confine excitons or drive their dynamics. Of course, any electrostatic potential has an opposite effect -- repulsive or attractive -- for the two particles of the pair carrying charge of opposite sign, and tends to dissociate the excitonic pair; hence, it competes with the attractive inter-particle interaction. A close cousin of this system are three particles complexes, with two like charges -- referred to as (negatively/positively) charged excitons or \textit{trions}  in semiconductor physics -- which is neither separable nor analytically soluble even in the absence of external potentials. Trions have been observed in a variety of systems, in view of their  prospects for device applications \cite{Datta2024}. Being weekly bound systems, trions are also a challenging playground for any theory, particularly in low-dimensions. 

 The multi-particle extension of the WMC method which we propose here is general, though, and we shall demonstrate its capability to simulate two and three particle systems bound by the mutual long-ranged Coulomb-like interaction, which we will often refer to  as excitons and trions, respectively, for brevity, borrowing the condensed matter nomenclature. However, we do not attempt to simulate specific physical systems, but rather focus on the fundamental and numerical aspects of the few-particle generalisation of the WMC approach.
	
	In this paper, then, we extend the WMC method to interacting particle systems which formally can handle an arbitrary number of particles, focusing our applications on Coulomb bound systems as discussed above, excitons and trions, generally in the presence of single-particle external potentials. We use the WMC approach to estimate the Feynman propagator of the multi-particle system, determining its late-time behaviour, and from this asymptotic regime we extract an estimate of the ground state energy. We systematically compare the WMC estimates with analytical values or a straightforward numerical diagonalisation of the (position space) Hamiltonian, whenever computationally possible. 
	
    The work presented here is meant to be a significant development of the WMC approach, both advancing the formalism and widening its scope. As such the algorithms presented below will be immediately adaptable to relativistic quantum systems and -- through the worldline framework -- to fresh numerical studies of quantum fields. Moreover, this article also explores computational aspects of the worldline algorithms that have not been explored in any depth in previous work. In particular we investigate the simulation time as a function of the number of worldline trajectories used in the numerical ensemble, examining its scaling behaviour in different dimensions and comparing this to the scaling dependence of the numerical diagonalisation algorithm. We find that the WMC approach scales more favourably than numerical diagonalisation with respect to both the number of discretisation points and the number of spatial dimensions of the quantum mechanical system.
		
	The article is organised as follows. In section \ref{sec:Theory} we outline the basic numerical technique for generating multiple Brownian motion trajectories (with and without external potentials). %We also present the theoretical background behind the exciton models %(section \ref{sec:Exc})
	Then, in section \ref{sec:Apps} we present our applications; first to some straightforward soluble systems and eventually, in section \ref{sec:nSep}, to the exciton system. 
	%Our numerical approach is compared to an approximate analytic description of exciton interactions in section \ref{sec:SE}. 
	Conclusions and our outlook are provided in \ref{sec:Concl}. Some technical details are given in the appendices: on our algorithms for generating trajectories in Appendix \ref{sec:ApAlg}, for dealing with singular potentials in Appendix \ref{sec:ApSmooth} and on computational complexity in Appendix \ref{sec:ApTimeScaling}.

	%%%%%%%%%%%%%%%%%%%%%%%%%%%%%%%%%%%%%%%%%%%%%%%%%%%%%%%%%%%%%%%%%%%%%%%%%%%%%%%%%%%%%%%
	\section{Background theory}
	\label{sec:Theory}
	
\subsection{Conventions and units}

To reduce the notational burden we take $\hbar = 1$ throughout. We will often name particles as \textit{electrons} and \textit{holes}, depending on the sign of their charge, and we label their particle trajectories -- without bold font -- by $x_{e} \in \mathbb{R}^{D}$ and $x_{h} \in \mathbb{R}^{D}$, respectively. We also take their masses to be equal as $m_{e} = m_{h} = 1$ (we shall comment on this choice of equal masses in the sections below). With respect to these units, we will also take the frequency of a harmonic oscillator interaction to be $\omega \sim \mathcal{O}(1)$ and fix the electric charge such that the inter-particle Coulomb interaction potential reads $V(r) = -\frac{\alpha}{r}$ with $\alpha = 1$ throughout, and $r$ the charged particles' separation. This will also affect our choices of electrostatic potentials providing attractive or repulsive potential wells (see section \ref{sec:softCoulombSqrWells}) and inter-exciton distances (see sections \ref{sec:IndirectExcitons}-\ref{sec:Trions}).

\subsection{Worldline numerics for multi-particle systems}
	\label{sec:Multi}
So far the WMC approach to simulations of the quantum mechanical path integral have been carried out according to equation \eqref{eq:PropWMC} only for single particle systems. This is the case for both non-relativistic quantum mechanics and for relativistic point particles\footnote{The reader is directed to \cite{TjonThesis, Nieuwenhuis:1996mc} for early Monte Carlo simulations of a simplified first quantised representation of a toy model scalar field theory using a thermalisation algorithm and to \cite{Zheng:2025} for studies of the Casimir effect for multi-atom systems, but still based on a constrained single particle trajectory.}. One can consider such simulations as a large mass / low velocity limit or mean field approximation of a multi-particle system where all but one of the particles is represented by an effective potential.

However, the natural question arises on the generalisation of the formalism for two or more particles, where correlation effects cannot be neglected. It is the aim of this section to describe the theoretical framework of such a generalisation; later sections will address the question of its scaling with particle number to consider whether the well-known efficiencies of the worldline approach are retained in multi-particle simulations. Although we restrict our application to quantum mechanics in this paper, the framework we advance here is immediately adaptable to relativistic point particles. Via the worldline formalism, it also applies to numerical simulations of quantum fields, allowing a long-delayed return to the line of research initiated in \cite{TjonThesis, Nieuwenhuis:1996mc, Savkli:2004bi, Nieuwenhuis:1995ux}.
	
Our aim is to describe a multi-particle propagator whose most general form is $K_{n}(\{x_{i}^{\prime}\}, \{x_{i}\}; T)$ with the path integral representation (in imaginary time formalism)
		\begin{equation}
			\hspace{-2mm} K_{n}=\int \mathscr{D}x_{1}(\tau_{1}) \cdots \int \mathscr{D}x_{n}(\tau_{n})\, \e^{- S[x_{1}(\tau_1),\ldots, x_{n}(\tau_n)]}\,, \label{eq:PropMultiGen}
		\end{equation}
		where we assume Dirichlet boundary conditions for each particle propagating from $x_{j}(0) = x_{j}$ to $x_{j}(T) = x'_{j}$, with $j=1,2,\ldots,n$.  The action reads
		\begin{align}
	\hspace{-1.0em}		S[x_{1}(\tau_1),\ldots, x_{n}(\tau_n)] &= \sum_{j=1}^{n}\int_{0}^{T} d\tau_{j} \,  \frac{m_{j}\dot{x}^{2}_{j}(\tau_{j})}{2} \nonumber \\
		\hspace{-1.0em}		&+ \prod_{j=1}^{n}\int_{0}^{T} d\tau_{j} \, V(x_{1}(\tau_{1}),\ldots, x_{n}(\tau_{n}))\,.
		\end{align}
In the non-relativistic setting which we are considering here, however, we will naturally focus on local interactions. Hence we work with a single time parameter $\tau$,	
		\begin{equation}
			\hspace{-2mm} K_{n}=\int \mathscr{D}x_{1}(\tau) \int \mathscr{D}x_{2}(\tau) \cdots \int \mathscr{D}x_{n}(\tau)\, \e^{- S[x_{1}(\tau),\ldots, x_{n}(\tau)]}\,, \label{eq:PropMulti}
		\end{equation}
		where, for an arbitrary potential that can incorporate local inter-particle interactions $V(x_{1}(\tau),\ldots, x_{n}(\tau))$, the action reduces to
		\begin{equation}
        \begin{split}
            \hspace{-0.75em}S[x_{1}(\tau),\ldots, x_{n}(\tau)] = \int_{0}^{T} d\tau \Big[ &\sum_{j=1}^{n} \frac{m_{j}\dot{x}^{2}_{j}(\tau)}{2}  \\ &+ V(x_{1}(\tau),\ldots, x_{n}(\tau))\Big]\,.
        \end{split}
\end{equation}
		 This path integral can be recast in terms of a properly normalised expectation value analogous to \eqref{eq:ExpectVal}, but with $n$ distinct path integrals for each trajectory with its corresponding Gaussian distribution on its velocity. Since for the free particle case ($V = 0$) the trajectories $x_{j}(\tau)$ are  independent, the normalisation factor will be a product of $n$ free-particle kernels $K_{0,j}$, and the multi-particle kernel reads
		\begin{equation}
			K_{n}(T)=K_{0,1}K_{0,2}\ldots K_{0,n} \Big\langle \e^{- \int_{0}^{T}d\tau \, V(x_{1}(\tau),\ldots, x_{n}(\tau))} \Big\rangle\,. \label{eq:PropMultiwExpVal}
		\end{equation} 
	where the endpoints of propagation for each free particle kernel are omitted for brevity,  and the expectation value over multi-particle trajectories is calculated (and normalised) with weight given by the product of free Boltzmann factors for each particle.		
	
	Equation (\ref{eq:Uasymp}) still holds for this multi-particle system, except that it is now described by a Hamiltonian \\$\Ham = \bigotimes_{j = 1}^{n} \Ham_{j}$ which generally takes the form
	\begin{align}
		\hspace{-1em}\Ham &= \Ham_{1}^{0}\otimes \id \otimes \ldots \otimes \id + \id \otimes \Ham_{2}^{0} \otimes \id \otimes \ldots \otimes \id + \ldots\nonumber \\
	\hspace{-1em}	& + \id\otimes \id \otimes \ldots \otimes \Ham_{n}^{0} 
		+ \Ham_{1}^{\I}\otimes \Ham_{2}^{\I} \otimes \ldots \otimes \Ham_{n}^{\I}\,,
	\end{align}
	where the $\Ham_{j}^{0}$ are the free particle Hamiltonians and the $\Ham_{j}^{\I}$ represent the interaction on each particle subspace of their tensor product of Hilbert spaces. For systems without a mass gap, an analogous version of the asymptotic functional form of the propagator, (\ref{eq:PropEspectralDecomp}), also holds with the wavefunctions $\psi^{(\star)}_{0}(x^{(\prime)})$ replaced by multi-particle generalisations $\Psi_{0}^{(\star)}(x_{1}^{(\prime)}, x_{2}^{(\prime)}, \ldots, x_{n}^{(\prime)})$,  but we shall discuss this technical point in more detail depending on the system under study below.
		
		Similarly to the single particle case, the main idea of the WMC formalism for multi-particle systems is to replace each continuous integral over trajectories in  \eqref{eq:PropMulti} by a finite sum $\int\mathscr{D}x_{j}(\tau) \to \frac{1}{N_{L}}\sum_{i=1}^{N_{L}}$ over $N_{L}$ paths, $\{x_{j,i}(\tau)\}_{i=1}^{N_{L}}$. For the purpose of evaluating the Riemann integral in the exponent we also introduce a discretisation along each of these trajectories in $\tau$, i.e., $x_{j,i}(\tau) \to \{ x_{j,i}(\tau_{k}) \}_{k=1}^{N_{p}}$. Furthermore, for a numerical implementation, we rescale the proper time to a dimensionless variable $u={\tau}/{T}$ and we expand each trajectory $x_{j}(u)$ about a straight line between the endpoints plus quantum fluctuations that are rescaled \textit{unit loops} $q(u) \in \mathbb{R}^{D}$, 
		\begin{equation}
			x_{j}(u)=x_{j}+(x'_{j}-x_{j})u + \sqrt{\frac{T}{m_{j}}}q_{j}(u)\,,\quad j=1,2,\ldots,n\,.
			\label{eq:xUnit}
		\end{equation}
		Now, the fluctuations must satisfy Dirichlet boundary conditions (DBC), $q_{0}=q_{N_{p}}=0$, and should have a Gaussian distribution on velocities so the joint probability density for a set of $n$ unit particle trajectories for a given loop, $i$, is
		\begin{align}
			\nonumber &\mathscr{P}_{n}\big( \{ q_{1,i},q_{2,i},\ldots,q_{n,i} \}\big) \sim \exp\left[{-\frac{1}{2}\int_{0}^{1}du\,\sum_{j=1}^{n}\dot{q}_{j,i}^{2}(u)}\right] \\
			\nonumber & \longrightarrow \exp\left[{-\frac{N_{p}}{2}\sum_{k=1}^{N_p}\sum_{j=1}^{n}(q_{j,i,k}-q_{j,i,k-1})^{2}} \right]\\
			& = \prod_{j=1}^{n} \exp\left[{-\frac{N_{p}}{2}\sum_{k=1}^{N_p}(q_{j,i,k}-q_{j,i,k-1})^{2}} \right]= \prod_{j=1}^{n} \mathscr{P}\big( \{ q_{j,i} \}\big)\,, \label{eq:GaussianDistMulti}
		\end{align}
        where $q_{j,i,k}=q_{j,i}(\frac{k}{N_p})$.
		Therefore, the expectation value is generalised to the following Monte Carlo estimation
		\begin{align}
			\nonumber &\left\langle \e^{-T\int_{0}^{1}du\,V(x_1(u),\ldots,x_{n}(u))}\right\rangle \\
			&\MC \frac{1}{N_L^n}\sum_{i_{1}, i_{2}, \ldots, i_{n}=1}^{N_L} \e^{-\frac{T}{Np}\sum_{k=1}^{N_p}V(x_{1,i_{1}}(u_{k}),\ldots, x_{n,i_{n}}(u_{k}))}\,. \label{eq:MCExpectValMulti}
		\end{align}
		Of course it is possible %-- and we have have found it convenient -- I found this not true (IVAN)
        to replace the expected value on the right side of (\ref{eq:MCExpectValMulti}) with the equivalent representation			
		\begin{align}
			 \frac{1}{\tilde{N}_L}\sum_{i=1}^{\tilde{N}_{L}} \e^{-\frac{T}{Np}\sum_{k=1}^{N_p}V(x_{1,i}(u_{k}),\ldots,x_{n,i}(u_{k}))}\,, \label{eq:MCExpectValMultiReduced}
		\end{align}
		where $\tilde{N}_{L}=N_L^n$ and $u_{k}=\frac{k}{N_{p}}$ is the discretisation of the rescaled proper time. This works because the $\{q_{j,i}\}$ are independently sampled from their Gaussian distribution on velocities (we have corroborated the equivalence in our simulations); however, we found that the numerical implementation with nested sums is faster than a single sum, so we implemented the former representation of the expectation value. Let us emphasise here that, as for the single-particle systems previously simulated using WMC \cite{UsMonteCarlo, Corradini:2020tgk}, we draw samples of (Brownian motion) trajectories according to the \textit{free} particle Gaussian distribution on velocities (this is why, of course, the probability distribution on trajectories factors in (\ref{eq:GaussianDistMulti})). We do this with the aim of retaining \textit{generality}, so that the framework so presented is in-principle applicable to arbitrary multi-particle systems, whilst recognising that there may well be problem-specific techniques that are more suitable for a particular physical system if the full information about the particle interactions can be incorporated into the sampling of trajectories. 
		
		Substitution of \eqref{eq:MCExpectValMulti} in \eqref{eq:PropMultiwExpVal}, along with the product of $n$ free particle propagators,
		\begin{equation}
		K_{0}(x',x;T) =	\Big(\frac{m}{2\pi T}\Big)^{\frac{D}{2}}\,\e^{-\frac{m}{2T}(x'-x)^{2}}\,,
		\end{equation}
		give us the worldline Monte Carlo estimation of the $n$-particle propagator,
		\begin{align}
		K_{n}(T) &=	\frac{\left(\prod_{j=1}^{n}m_{j}\right)^{\frac{D}{2}}}{(2\pi T)^{\frac{D}{2}n}}\,\e^{-\sum_{j=1}^{n}\frac{m_{j}}{2T}(x'_{j}-x_{j})^{2}}\, \nonumber \\
		&\times \frac{1}{N_L^n}\sum_{i_{1}, i_{2}, \ldots, i_{n}=1}^{N_L} \e^{-\frac{T}{Np}\sum_{k=1}^{N_p}V(x_{1,i_{1}}(u_{k}),\ldots, x_{n,i_{n}}(u_{k}))}\,. \label{eq:WMCfull_n-Propagator}
		\end{align}
		In Appendix \ref{sec:ApAlg} we present the adaptation of an algorithm to generate Brownian motion trajectories according to the measure (\ref{eq:GaussianDistMulti}), now for \textit{multiple} particles. Let us note here that the computational implementation of this algorithm is, by construction, rejection free (and the independent generation of trajectories ensures no auto-correlation); also, for the \textit{local} interactions considered here we can evaluate both sums in \eqref{eq:WMCfull_n-Propagator} without storing the spatial points in memory.
		
		Let us mention in passing how this compares to approaches based on sampling trajectories according to the full action (that takes the interaction potential into account). There the aim is to determine the expectation value of operators (e.g.~$\hat{x}$, $\hat{\pp}$ or more generally $\hat{\mathcal{O}}(\hat{x}, \hat{\pp})$) over trajectories $\{x(\tau)\}$ weighted by the Boltzmann factor $\e^{-S[x(\tau)]}$. Denoting such expectation values by 
		\begin{equation}
			\llangle \hat{\mathcal{O}} \rrangle := \frac{\int \mathscr{D}x(\tau) \, \mathcal{O}\big(x(\tau), m\dot{x}(\tau)\big) \, \e^{-S[x(\tau)]}}{\int \mathscr{D}x(\tau) \, \e^{-S[x(\tau)]}}\,,
		\end{equation}
		it is not usually stated how such expectation values can be obtained in the WMC approach based on free-particle trajectories. A simple relation exists that allows them to be estimated using WMC:
		\begin{equation}
			\llangle \hat{\mathcal{O}} \rrangle = \frac{\langle \big(\mathcal{O}(x(\tau), m\dot{x}(\tau)\big) \, \e^{-\int_{0}^{T} d\tau V(x(\tau))} \rangle }{\langle \e^{-\int_{0}^{T} d\tau V(x(\tau))} \rangle}\,,
		\end{equation}
		which simply re-weights the Brownian motion samples of the operator with the Wilson line evaluated on the trajectory from which the sample was obtained.

		Naturally, when moving from the continuous functional integral to a finite discretisation of the path integral, we introduce errors that must be addressed. In the WMC formalism, there are two main sources of such errors. The first one is a systematic error associated to the discretisation of the worldlines with respect to their parametrisation, $u$, which is difficult to estimate when we calculate the Wilson line integral. So far in the the literature, there is no mathematical expression to estimate the systematic error, but it has been investigated in \cite{UsMonteCarlo} by studying the convergence of the path integral estimation with respect to $N_{P}$. There is widespread evidence that $N_p \sim 10^3\;\text{to}\;10^4$ provides good estimations for one particle simulations \cite{Gies:2000zp,Gies:2001tj,UsMonteCarlo,Franchino-Vinas:2019udt,Corradini:2020tgk}; here we will use the same order of number of points per particle. 
        
        The second source of error is a statistical error due to replacing the path integral(s) with a finite sum over trajectories. Here, the standard error of the mean across the total number of loops, $\tilde{N}_{L}$, provides a faithful estimate, whose generalisation to the case of many particles is described as follows. Letting $W_i$ be the Wilson line value for the $i$-th loop, the standard error of the mean reads
		\begin{equation}
			\text{SEM}(W)=\sqrt{\sum_{i=1}^{\tilde{N}_{L}}\frac{(W_i-\langle W \rangle)^2}{\tilde{N}_{L}(\tilde{N}_{L}-1)}}=\sqrt{\frac{\sum_{i=1}^{\tilde{N}_{L}}W_i^2-\tilde{N}_{L}\langle W \rangle^2}{\tilde{N}_{L}(\tilde{N}_{L}-1)}}, \label{eq:SEM}
		\end{equation}
		where $\langle W \rangle=\sum_{i} W_i/\tilde{N}_{L}$. In the final expression we have, of course, just expanded the rounded bracket of the variance, but this latter expression has numerical advantages with respect to a finite machine precision. Furthermore, to calculate the SEM it is not necessary to store all the trajectories, only to calculate their individual contribution to the Wilson line and to its expected value.
		
	This concludes the presentation of our numerical approach. The remainder of this manuscript is concerned with implementing the numerical simulation and testing its characteristics as an estimation of the full multi-particle propagator. Our goals include determining how accurate the numerical results are; how the simulations scale with $\tilde{N}_{L}$ (and suitable values to ensure accurate results) and the processing power required for robust estimations. We also corroborate the validity of this multi-particle generalisation of the WMC formalism by studying a simple model of excitons, including the presence of external potentials, and trions. In the following section, we describe these excitons in detail before going on to estimate their binding energies and behaviour when exposed to external potentials.

	%%%%%%%%%%%%%%%%%%%%%%%%%%%%%%%%%%%%%%%%%%%%%%%%%%%%%%%%%%%%%%%%%%%%%%%%%%%%%%%%%%%%%%%%%
	\section{Applications}
	\label{sec:Apps}
    
	In this section we shall first demonstrate the reliability and effectiveness of the proposed multi-particle method with two-particle systems with internal interactions only (which we often call a free exciton), which therefore can in principle be separated and reduced to a one-particle problem (and, in some case, it can also  be solved analytically). Then we progress to systems with both inter-particle interactions and externally applied potentials, which can only be solved exactly (i.e. to arbitrary precision) by numerical techniques. In each case, we extract the ground state energy of the system from the late-time behaviour of the propagator (as described below) and compare the result to  either analytical or to numerical diagonalisation results where possible.
	
	\subsection{Separable systems}
	\label{sec:Sep}

		As a first benchmark of the method, we consider a two-particle model with an attractive inter-particle interaction only (free exciton), that depends on the separation of the particles' positions, $V(x_{1}(u),x_{2}(u))=V(|x_{1}(u)-x_{2}(u)|)$. Of course, for such translationally invariant interactions the centre of mass (CM) and relative DoFs separate, with the CM behaving as a free particle with mass $M = m_{1} + m_{2}$, and the two-particle problem is reducible to a trivial one-particle problem in the relative coordinate, which can be solved analytically, for a suitably chosen potential, or numerically with increased efficiency. Since we are interested in setting up multi-particle simulations for the most general case (see Sec.~\ref{sec:nSep}), however,  here we explicitly treat both particles' DoFs also for separable systems, thereby retaining the full CM dynamics within the WMC framework. The inclusion of the CM DoFs require a careful treatment in the estimation of ground state energies, though, as the contribution from the continuum of eigenstates of the Hamiltonian must be properly accounted for.
			
	\subsubsection{Harmonic oscillator interaction}
	\label{sec:HO}

    To validate our method, we consider first an electron-hole pair bound by an harmonic oscillator potential. This choice makes for an analytically tractable model, and is used as a test of accuracy and performance, regardless of its physical significance. We study the system in both one and two spatial dimensions, comparing the ground state energies so estimated with exact results. 

    The interaction potential for this model reads
	\begin{equation}
		V(|x_e(u)-x_h(u)|) = \frac{1}{2}\mu\omega^{2}|x_{e}(u) - x_{h}(u)|^2,\; x_{e}, x_{h} \in \mathbb{R}^{3}\,,
	\end{equation}
	where the reduced mass of the two-particle system $\mu=m_{e}m_{h}/(m_{e}+m_{h})$, and  $\omega$ is the angular frequency of the harmonic interaction. Later in the paper we shall turn to Coulomb bound systems, where the Coulomb singularity is removed by a parameter $d$ (which can be thought of as the minimum physical separation between the electron and the hole -- see Sec. \ref{sec:IndirectExcitons}). In the same spirit, we  assume a minimum particle separation $d$ in the harmonic potential. This constant introduces a tunable offset that shifts the potential minimum and allows us to dimensionally reduce the system to one with potential 
	\begin{equation}
		V(|x_e(u)-x_h(u)|) = \frac{1}{2}\mu\omega^{2}(|x_{e}(u) - x_{h}(u)|^2 + d^{2})\,,
	\end{equation}
	with now $ x_{e}, x_{h} \in \mathbb{R}^{1}$ or $\mathbb{R}^{2}$ depending on the dimensionality $D$ of the system.

	It is, of course, a textbook exercise to solve this system by turning to CM and relative coordinates, $R(\tau) = \frac{1}{2}(x_{e}(u) + x_{h}(u)) $ and  $r(u) = x_{e}(u) - x_{h}(u)$, respectively, and momentum operators, $\widehat{\PP}$ and $\hat{\pp}$,  in which $\Ham = \Ham_{R}\otimes \id + \id \otimes \Ham_{r}$ with
	\begin{align}
		\Ham_{R} &= \frac{\widehat{\PP}^{2}}{2M}\,, \\
		\Ham_{r} &= \frac{\widehat{\pp}^{2}}{2\mu} + \frac{1}{2}\mu\omega^{2}(\widehat{r}^2 + d^{2})\,.
	\end{align}
 where $M = m_e + m_h$  and $\mu = \frac{m_e m_h}{m_e + m_h}$ are the total and reduced masses. The eigenstates and eigenvalues can be immediately deduced, with energies being
	\begin{equation}
		E(P, n) = \frac{P^{2}}{2M} + \omega\Big(\frac{D}{2} + n\Big) + \frac{1}{2}\mu\omega^{2}d^{2}\,.
	\end{equation}
	In the limit of low CM momentum the spectrum is essentially controlled by the relative coordinate which has ground state 
	\begin{equation}
	E_{r, 0} = \frac{\omega}{2}(D + \mu\omega d^2)\,.
	\label{eq:groundEnergyHO}
	\end{equation}
	Note, however, there is a continuum of energies associated to free particle motion of the CM.

	 For this model, the propagator can be calculated analytically,  reading
	\begin{align}
		\hspace{-0.25em}	K = \e^{-\frac{1}{2}\mu\omega^2 d^2 T} \left(\frac{M}{2\pi T}\right)^{\frac{D}{2}} 	\e^{-\frac{M(R'-R)^2}{2T}} \left(\frac{\mu\omega}{2\pi\sinh{(\omega T)}}\right)^{\frac{D}{2}} \nonumber  \\
		\hspace{-0.25em}	\exp{\left(-\frac{\mu\omega}{2\sinh{(\omega T)}}\Big[(r'^{2}+r^{2})\cosh{(\omega T)}-2r'r\Big]\right)},
		\label{eq:propHO}
	\end{align}
	where the constant shift in the energy spectrum emerges in the imaginary time formalism as an exponential damping factor.	 
	
	Following the formalism represented by equation \eqref{eq:PropMultiwExpVal}, the WMC representation of the propagator for this system becomes
	\begin{align}
	  %\hspace{-1.25em}	
      &K(x_e^{\prime},x_h^{\prime}, x_e,x_h;T)=K_{0,e}K_{0,h}\nonumber\\&\quad\quad\times\Big\langle \e^{- \frac{1}{2}\mu\omega^2T\int_{0}^{1}du \, (|x_{e}(u) - x_{h}(u)|^2 + d^2)} \Big\rangle
		\label{eq:WMCpropHO}  
	\end{align}
	where $K_{0,e}$ and $K_{0,h}$ are the free-particle propagators for the electron and the hole respectively, and the expectation value is taken over all pairs of oriented worldlines in the path integral ensemble that join their respective endpoints. We emphasise, again, that we do \textit{not} wish to reduce this path integral to one only over the separation of the trajectories -- we retain the full dynamics by simulating the two particle trajectories separately as proof of concept for the multi-particle WMC simulations developed here. Let us also add that the dependence on the parameter $d$ is simply a multiplicative factor in $K$ (in the real time formalism it would be pure phase), we nonetheless retain this since it allows us to test our procedure for estimating the ground state energy of the system (and recover the quadratic dependence of the energy in \eqref{eq:groundEnergyHO}). This is important as a benchmark of our method, given that prior work (e.g. \cite{UsMonteCarlo, UsPv}) found that the WMC simulations can be numerically sensitive to the parameters such as mass or coupling strength chosen.
            	
	\begin{figure}
		\centering
		\includegraphics[height=0.275\textheight]{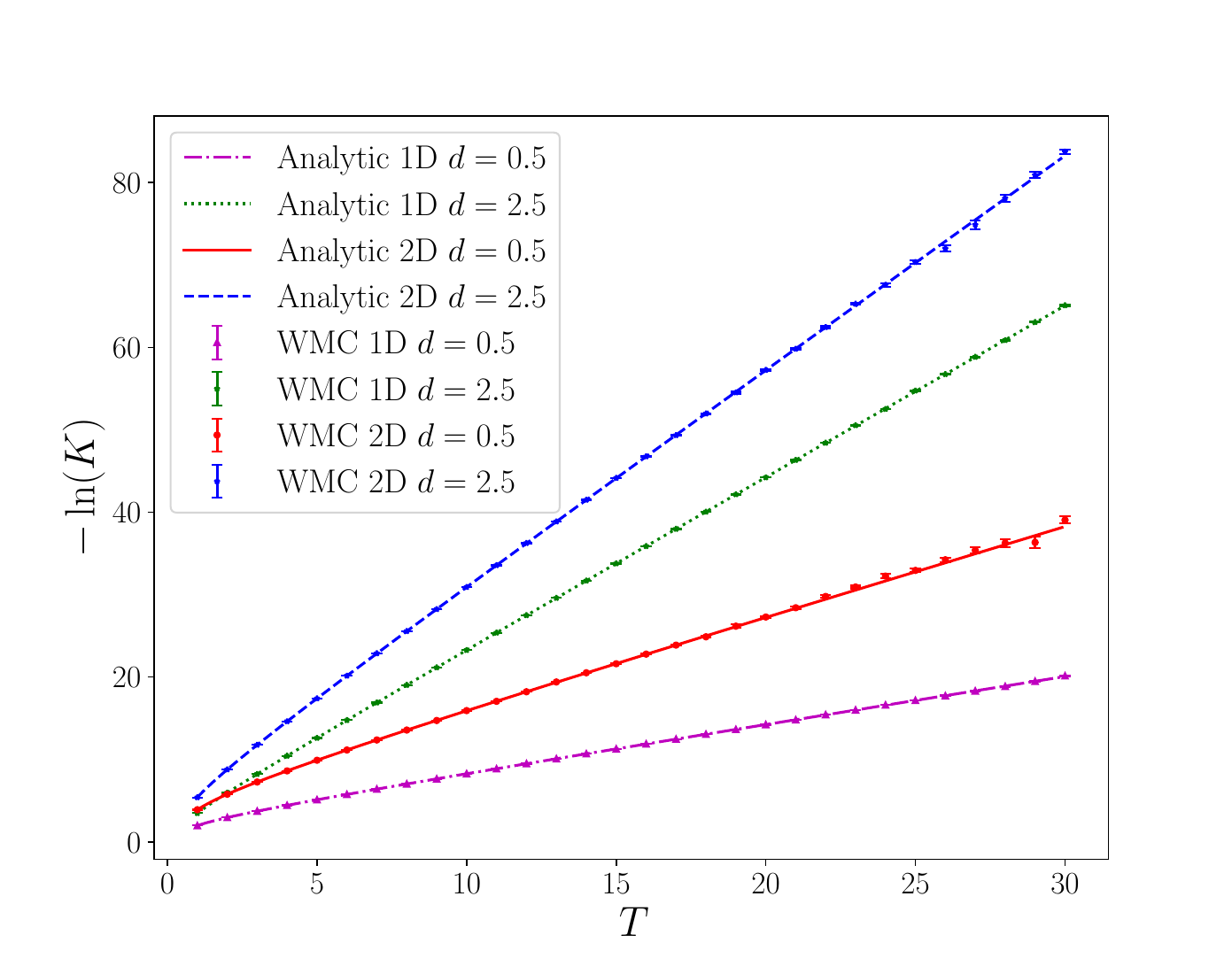}
		\caption{WMC estimation for the logarithm of the kernel for two particles with harmonic interaction in 1D and 2D for two values of the parameter $d$ (see equation~\ref{eq:propHO}). Calculations performed with $N_L=1000^2$ , $N_p=5000$,  $\omega=1$ . Dots: WMC estimates with statistical error bars. Lines: analytic results (Eq.~\ref{eq:propHO}).}
		\label{Fig:HO_lnK}
	\end{figure}
    
	Figure~\ref{Fig:HO_lnK} demonstrates good agreement between the WMC estimation of (the logarithm of) the propagator, simulating \eqref{eq:WMCpropHO} using the algorithms in Appendix \ref{sec:ApAlg}, and the analytical result in \eqref{eq:propHO}. The parameters used for this simulation are given in the figure and we postpone the discussion of  simulation time and complexity to a later section. The error bars displayed in the figure were calculated using \eqref{eq:SEM} and the first order error in the logarithm of the kernel, $\Delta(\ln(K)) \approx \Delta K/K$. All data points are within one standard deviation of the known analytic result.
	
	To compare the ground state energies obtained via the WMC formalism with the analytic result, (\ref{eq:groundEnergyHO}), one would normally use the natural logarithm of \eqref{eq:PropEspectralDecomp}, suitably Wick rotated and extended to the multi-particle case, $\lim\limits_{T \to \infty} \ln(K(\{x'\}, \{x\}; T)) \sim \ln(\Psi_{0}(\{x'\})\Psi_{0}^{\star}(\{x\})) -E_{0}T$. Then, $E_0$ can be extracted from the following asymptotic limit
				\begin{equation}
					E_0 = -\lim_{T \to \infty}\frac{d}{dT}\ln(K(\{x'\}, \{x\}; T ).
				\end{equation}
	This means that we can perform a linear fit in some time window  where the logarithm of the propagator is linear, and take the gradient of such a fit to be the numerical estimation of the ground state energy --for more details in prior literature see \cite{UsMonteCarlo,UsPv}. This apparently seems reasonable given a first glance at Fig.~\ref{Fig:HO_lnK}.	Nevertheless, if we want to apply this procedure to separable systems, we must take into account the lack of mass gap due to the continuous spectrum from the CM DoFs. Hence, to compare WMC results with the ground state energy (only associated to the relative separation), we must adapt the asymptotic fit so as to correctly accommodate the CM contribution.
		
	To illustrate this mechanism, we consider the total wavefunction of a separable system (e.g the harmonic oscillator) written as a product of CM and relative coordinates:
	$\Psi(R,r)= \phi(R)\psi(r) $. 
	To isolate the relative (internal) dynamics, we integrate over the momentum contribution from the CM motion in the imaginary time propagator:
		\begin{align}
		K(\{R', r'\}, \{R, r\}; T) &= \sum_{n=0}^{\infty} \, \psi_{n}(r')\psi_{n}^{\star}(r) \e^{-E_{n}T}\nonumber \\
		& \times \int d^{D}P \,  \phi(R')\phi^{\star}(R) \, \e^{- \big(\frac{P^2}{2M} + \ldots\big) T} \,,
	\end{align}
	where we allow for non-trivial dynamics for the CM system, asking only that it has a continuous spectrum (for the free particle, of course, $\sqrt{2\pi} \phi(R) = \e^{i P \cdot R}$). In the large $T$ limit, the exponential damping from excited states with energies $E_{n} \gtrsim 1/T$ and $P^{2} \gtrsim 2M/T$ makes these contributions subleading. For the discrete spectrum of the relative coordinate, this projects immediately onto the ground state. For the continuum of energies of the CM system, we cannot separate any lowest-energy state from the rest of the spectrum. However, integrating within this region gives the asymptotic form
	\begin{equation}
	\hspace{-0.25em}	K(\{R', r'\}, \{R, r\}; T) =  \psi_{0}(r')\psi_{0}^{\star}(r) \e^{-E_{0}T} \times T^{-\frac{D}{2}}f(R', R; T)
	\end{equation}
	where $f(R', R; T) \sim h(R', R)/T + \mathrm{const.}$ at large $T$. Hence taking the logarithm gives a linear-plus-logarithmic expression as the leading contributions in the large time limit:
		\begin{align}
			-\ln K(T) = -a + E_0 T + b \log T\,,
			\label{eq:non-linear.fit}
		\end{align}
	where $a$ is an irrelevant shift coming from the ground state wavefunction evaluated at the endpoints and some normalisation constants and $b$ is a parameter that depends on the dimensionality. This expression correctly separates the contribution from the CM and the ground state energy given by the relative motion and, therefore, will be the one we use to estimate the (reduced) ground state energy of separable systems. In the particular case of the harmonic interaction outlined above,  this asymptotic form can be obtained exactly in the large $T$ limit of equation ~(\ref{eq:propHO}) and indicates that there are subleading corrections in $\log(T)$ to the lines in Fig.~\ref{Fig:HO_lnK}.
	
	To apply equation~ \eqref{eq:non-linear.fit} to estimate the ground state energy $E_0$, using the numerical estimation of the imaginary time propagator from \eqref{eq:WMCpropHO}, we have to determine a compatibility window for a range of distances $d$ where $-\ln(K)$ follows the asymptotic form in equation ~(\ref{eq:non-linear.fit}) (typically about 15 units of time $T$). Then we run simulations in this window using an independent set of trajectories for each value $T$ to avoid autocorrelations and we average over a suitable number of repetitions\footnote{We used 100 repetitions for each value of $T$ in all of our simulations for estimations of $E_0$, typically with a separation $\Delta T = 0.1$.} to reduce the statistical error in the simulations, as demonstrated in \cite{UsMonteCarlo}. This procedure is computationally expensive; however one can exploit the natural parallel nature of the WMC formalism \cite{Aehlig:2011xg,Mazur:2014hta} to distribute the simulations across multiple cores on a HPC (more information is given in Appendix \ref{sec:ApAlg}). 
    Finally, we fit equation~ \eqref{eq:non-linear.fit} with a weighted nonlinear fit using the Mathematica function \textsf{NonlinearModelFit}  \cite{Mathematica}, taking the weights to be given by the corresponding values of $1/[\Delta(\ln(K))]^{2}$ (i.e. estimations with smaller error have greater weight)\cite{Strutz:2016}. The error in each estimate of $E_0$ was then taken to be the error associated to the weighted fit.
		
		\begin{figure}
			\centering
			\includegraphics[height=0.275\textheight]{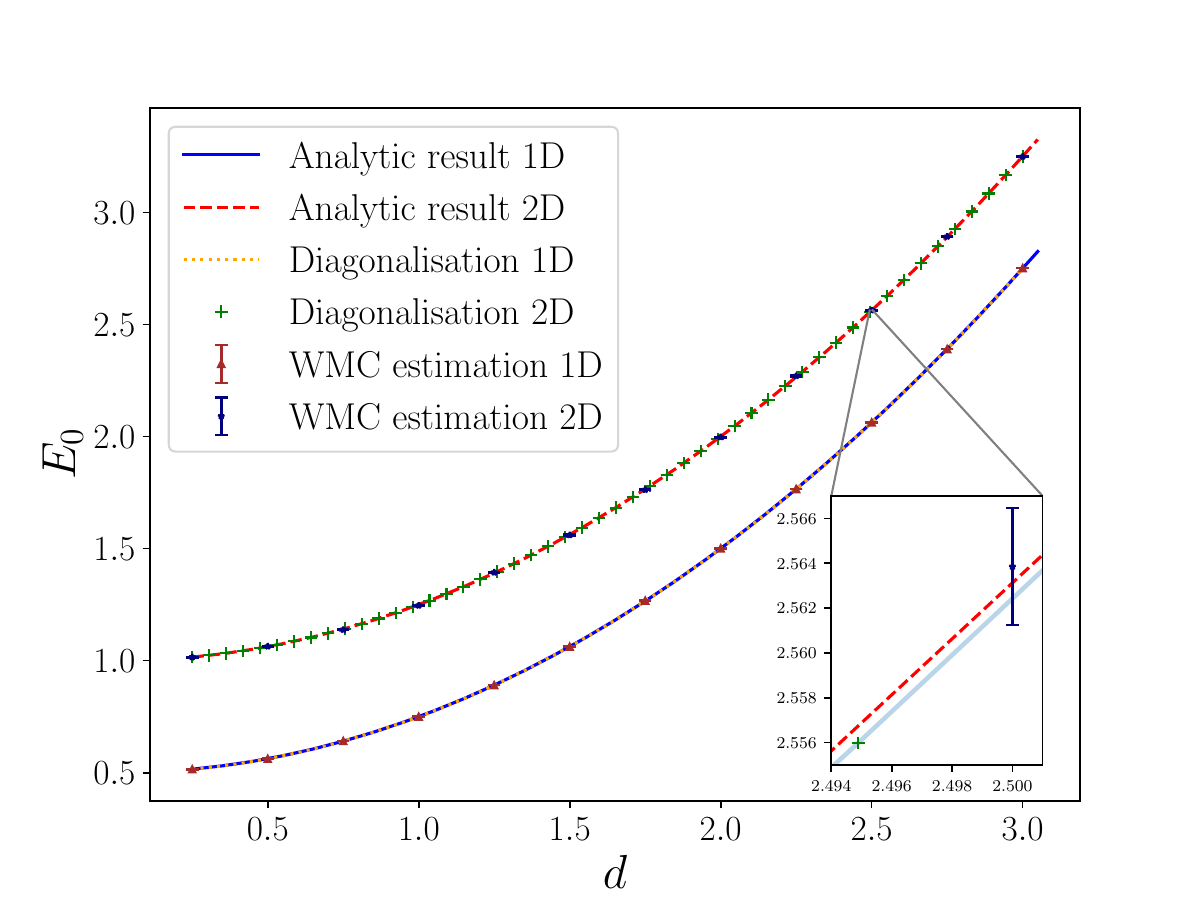}
			\caption{Ground state energy estimates for two particles in 1D and 2D with harmonic oscillator interaction with  $\omega=1$ . Parameters of the WMC calculation are $N_L=500^2$ , $N_p=10000$.  Parameters for diagonalisation: $N_{T}=300$,  $N_{T}=300^{2}$  in 1D,  2D, respectively. Inset: zoom at $d=0.25$ in 2D.}
			\label{Fig:HO_groundEnergy}
		\end{figure}

	In Fig. \ref{Fig:HO_groundEnergy}, the results from our simulations show good agreement with the analytic predictions of equation \eqref{eq:groundEnergyHO} for one and two dimensions, with parameters $\omega=1$ and $\mu=\frac{1}{2}$. The errors in the
	estimates were found to be of order of magnitude $\mathcal{O}(10^{-3})$ and our predictions are within 1 standard error from the analytic result -- see the inset figure providing a clearer display of the size of the numerical error (that is representative of points throughout the plot). Note that this would not have been the case had we instead made a \textit{linear} fit to the late time form of $\log(K)$, so ignoring the logarithmic corrections arising from the CM spectrum.

    \subsubsection{Comparison to a standard numerical approach}

      For the remaining models under study below, solutions in closed form are generally not available. Hence, our WMC results will be systematically compared with ones obtained by solving the Schr\"odinger equation for the ground state using a straightforward numerical diagonalisation procedure, which we outline here:  we discretise space to a grid of points in $D$ dimensions, with a number of points, $N$, evenly spaced in a certain interval (per particle per dimension), such that the total number of grid points is given by $N_{T}=N^{nD}$. On this grid we evaluate a finite dimensional approximation to the Hamiltonian as a large, sparse matrix, and we apply the Implicitly Restarted Arnoldi Method to compute its eigenvalues using the ARPACK library~\cite{lehoucq1998arpack}.  Errors in this ``diagonalisation method'' are estimated by extracting an approximation to the true value of the ground state energy, or lowest eigenvalue, by extrapolating to an infinite grid resolution ($N \to \infty$, the continuum limit). The error in the ground state energy for a given number of grid points, $N$, is estimated as the difference from this asymptotic value $E_{0,\infty}$, that is,
      \begin{equation}
        \Delta E_{0,N} \approx |E_{0,N} - E_{0,\infty}|\,.
      \end{equation}
      Here $\Delta E_{0,N}$ is the error of the ground state energy estimated for a specific number of grid points $N$ and $E_{0,N}$ is the value calculated from the simulation.
      
      To calculate the asymptotic value $E_{0,\infty}$, we take several estimates of $E_0$ with different numbers of grid points $N$ for a particular value of $d$. We fit a Pad\'e approximant as a function of the inverse of the number of grid points, i.e., $R(1/N)$, of order $m=n$ with $m=n=3,4$. We plot the fits given by the Pad\'e approximations against the data to see how well they capture the trend, and finally, we take the $N\to\infty$ limit, so the asymptotic value is only the first coefficient of the Pad\'e approximant fit\footnote{This choice of asymptotic fit was partly motivated by \cite{nHit}, where a Pad\'e approximant was used to improve WMC estimates of the propagator with highly localised potentials. Alternatives exist, of course, such as Richardson extrapolation \cite{Richardson}, Aitken diagonalisation \cite{Aitken} etc -- the convergence of the Pad\'e approximant could also be accelerated using the Shanks transformation \cite{Shanks} as in \cite{nHit}. Our intent here, however, is just to get a simple order of magnitude estimation of the numerical error.}.
      
      This is a time-consuming process if done for each $d$, in the simulations, so for each system (in 1D or 2D) we performed the above process for points $d \in \{0.25,1.0,2.75\}$. The error for each point $d \in [0,3]$ is then estimated as $\Delta E_{0,N}(d)=a+bd+cd^{2}$ with the coefficients $a$, $b$ and $c$ calculated by solving the system of equations for the error given the previous three points; in this way, we can obtain a reasonable order of magnitude estimation of the error for this numerical approach. This approach has been validated against the analytical results for the harmonic oscillator interaction, as shown in Fig.~\ref{Fig:HO_groundEnergy}.
      
      In Appendix \ref{sec:ApTimeScaling} we show how the diagonalisation method scales in time compared to the WMC formalism for calculating the ground state energy and we study the convergence of the diagonalisation method for external potentials, where the Pad\'e approximants are shown explicitly. We demonstrate that the WMC formalism scales linearly in time and more favourably than the diagonalisation method, except in 1D.
            	
	\subsubsection{Soft Coulomb interaction}
	\label{sec:IndirectExcitons}
	
We now turn to electron-hole systems interacting via the Coulomb potential. In free space, after transformation into CM and relative coordinates,  the wavefunction separates in $\Psi(R, r) = \phi(R) \psi(r)$ and the Hamiltonian for a spinless electron-hole pair reads
		\begin{equation}
			\Ham = \frac{\widehat{\PP}^2_{R}}{2M} + \frac{\widehat{\PP}^2_{r}}{2\mu} - \frac{e^2}{4\pi \epsilon_0 \epsilon_r |\widehat{r}|}\,,
            \label{eq:Coulomb3D}
		\end{equation}
which leads to a bound, Coulomb  correlated ground state. Note that we take equal electron and hole masses, although they may differ, for example, for electrons and holes in condensed matter systems. This choice limits the number of parameters. However, there is the more subtle advantage of maximising correlation effects in the ground state of the system; hence, it is a strict test for the new WMC method. We shall return to this point when treating non-separable systems, in Sec.~\ref{sec:nSep}.
 
The Coulomb potential in equation.~\eqref{eq:Coulomb3D}  must be understood as the electron-hole interaction in 3D. It has a troublesome singularity at the origin which, in lower dimensionality, would cause, for example, the divergence of the exciton binding energy in 1D. Actually, when reducing the dimensionality to 2D or 1D one should be careful about quantum confinement effects: for example, with strong confinement in a given direction $\eta$ the 3D ground state wavefunction can be accurately factorised as  
        \begin{equation}
			\Psi_{3D}(r) = \gamma(\eta)\psi_{2D}(r_{\parallel})\,,
			\label{eq:separable-wf}
		\end{equation}
which neglects contributions from higher, well separated quantum confined states along $\eta$. The 2D effective interaction between particles separated by $r_\parallel$  should be averaged over  $\gamma(\eta)$, which results in a finite value at $r_\parallel=0$, while maintain the long-range Coulomb tail.

Derived effective 1D or 2D Coulomb interactions from transverse confinement depend on the dimensionality and the form of $\gamma(\eta)$, but they can be generally approximated with accuracy by a soft-core regularised form \cite{PhysRevB.68.045328} 
		\begin{equation}
			V_{\text{SC}}(r)=-\frac{\alpha}{\sqrt{r_{\parallel}^2 +d^2}},
			\label{eq:Soft-Coulomb}
		\end{equation}
where $\alpha=\frac{e^2}{4\pi \epsilon_0 \epsilon_r}$. The number of spatial dimensions and details of the confinement, which determine  $\gamma(\eta)$, are abstracted into $d$, which turns out to be of the order of the confinement length along $\eta$. In a different perspective, in condensed matter nano-systems, excitons may also form from electron-hole pairs with the two particles sitting in separated, parallel quantum confined layers or wires -- aka spatially indirect excitons -- where $d$ now represents the average distance between the two particles.  Aside from being related to the physics of the system,  the parameter $d$ has the added benefit of softening the hard singularity when $|r_{\parallel}|\to 0$, playing a similar role to the cut-off parameter used in \cite{RLoudon1959Hydrogen1D,RLoudon2016Hydrogen1D}. 

The interaction in equation \eqref{eq:Soft-Coulomb} is a challenging playground to test the multi-particle WMC method. On the one hand, it is long-ranged. On the other hand, the regularisation parameter $d$ allows us to test different regimes, as $d$ determines the localisation length of the pair, with the pair experiencing different regions of the potential. Before embarking in a systematic application of the multi-particle WMC to interaction in equation \eqref{eq:Soft-Coulomb}, however, we note that for the limit case $d=0$ analytical solutions are well known: the ground state energy $E_0 \rightarrow -\infty, E_0=-4 Ry$ for 1D, 2D, respectively, where $Ry$ is the effective Rydberg ($Ry= 0.25$ in our units) coinciding with the 3D binding energy. From the numerical point of view, however, since the potential becomes singular at $r_{\parallel} = 0$,  a special regularisation -- ``\textit{smoothing}'' -- is needed in the WMC formalism (this is outlined in Appendix \ref{sec:ApSmooth}, where we show that the smoothing technique developed in \cite{UsMonteCarlo} can be lifted to the multi-particle case). On the other hand, the exact analytic solution to this quantum mechanical system is not known in closed form for $d \neq 0$ \cite{Knox, Comb}, in any dimension of space due to the added complication of the softening parameter, $d$ (but see \cite{SoftAnal, SoftAnal2} for approaches based on infinite series). However, numerically exact solutions are easily obtained, and variational estimations are available \cite{Grasselli2017VariatinalSoftCoulomb}.

		The WMC estimation of the propagator for the dimensionally reduced system is given in terms of particle trajectories, $x_{e}$ and $x_{h}$, associated to the electron and hole by  
			\begin{equation}
                \begin{split}
              	K(x_e^{\prime},x_h^{\prime}, x_e,x_h;T) &=   K_{0,e}K_{0,h} \\ 
                &\times \Big\langle \e^{ -T\int_{0}^{1}du \, V_{\text{SC}}(|x_{e}(u)-x_{h}(u)|)} \Big\rangle\,,
                %\Big\langle \e^{ -T\int_{0}^{1}du \, \frac{1}{\sqrt{|x_{e}(u)-x_{h}(u)|^2 +d^2}}} \Big\rangle\,,
				\label{eq:WMCpropFreeExciton}
                \end{split}
			\end{equation}
		which is ready to be simulated using the WMC algorithm outlined above. Note, again, that in the numerical systems we do not work with CM and relative coordinates here, instead choosing to simulate the full DoFs. We then apply (\ref{eq:non-linear.fit}) to estimate the binding energy of the exciton. 
        
\begin{figure}[t]
		\centering
		\includegraphics[width=0.45\textwidth]{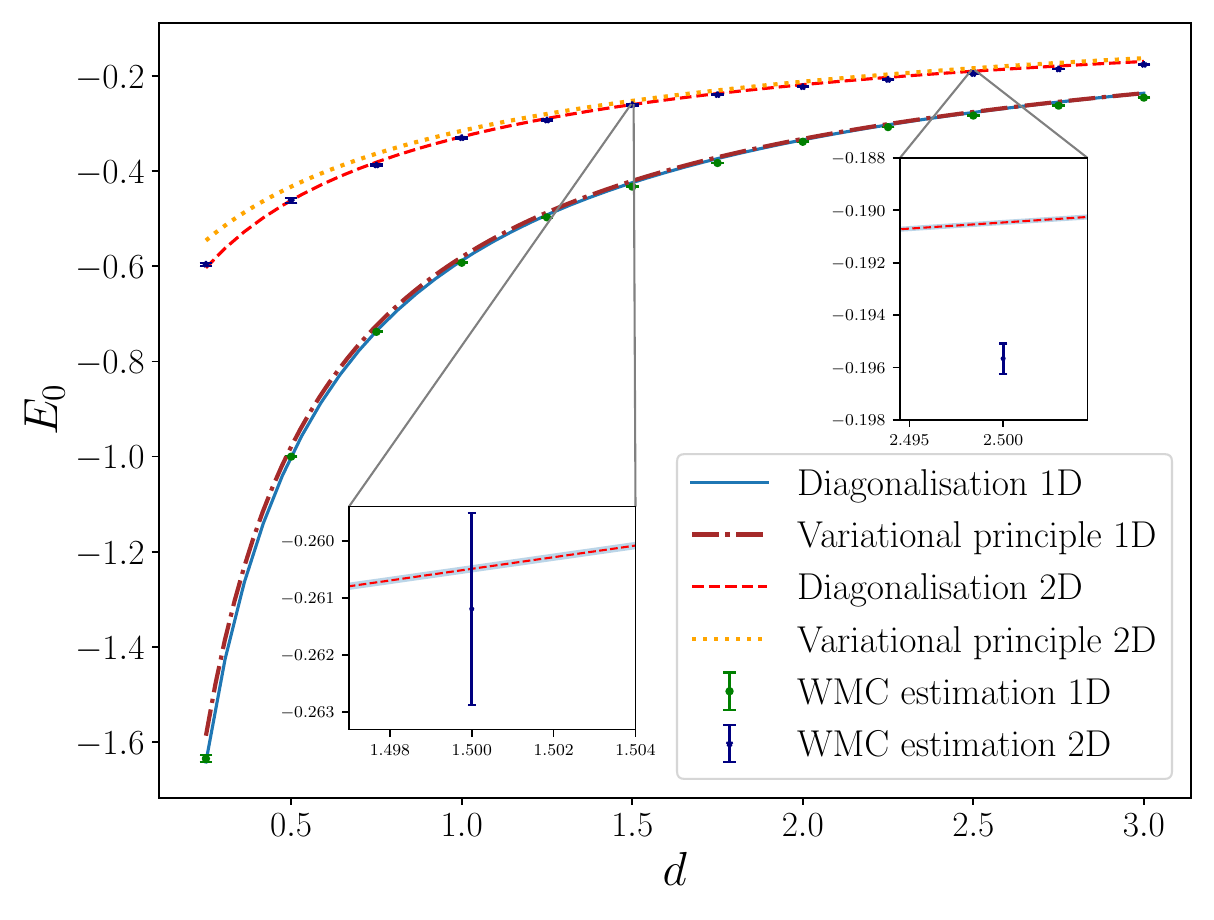}
		\caption{Ground state energy  of two particles in 1D and 2D interacting via a soft-Coulomb interaction vs the cut-off parameter $d$. For WMC: $N_L=500^2$ , $N_p=10000$.  For diagonalisation: $N_{T}=300$,  $N_{T}=300^{2}$  in 1D, 2D, respectively. Insets: zoom at $d=1.5\;\&\;d=2.5$ in 2D.}
		\label{Fig:softCou_groundEnergy}
	\end{figure}

        In Fig.~\ref{Fig:softCou_groundEnergy}  we compare the ground state energy of the system in 1D and 2D obtained from the WMC calculation with numerical estimates provided by the diagonalisation method and analytical estimates obtained using variational Gaussian trial wavefunctions, as developed in \cite{Grasselli2017VariatinalSoftCoulomb}.
		
		We find an overall good agreement between our WMC calculations and numerical diagonalisation in the investigated range $d \in [0.25,3]$, with almost all data points being within the numerical error, except  for the largest values of $d$, where  a tiny disagreement between the numerical approaches is found (see representative insets). Importantly, the agreement is similarly good for both 1D and 2D, which have very different binding energies, with 1D binding being stronger, as is well known \cite{Davies:1997}.  The variational principle, being an upper bound on the ground state energy, confirms the quality of the numerical calculation. This worsens for $d<1$ in 2D (to a lesser extent in 1D)\footnote{This is expected considering that, for small $d$, the Coulomb interaction starts to dominate and the dynamics of this region can not be well described by a harmonic oscillator type trial wavefunction; remarkably, though, it reproduces the correct behaviour of the system for $d>1$ in both dimensions.}. Overall, we conclude that the WMC methods captures well the internal relative dynamic which is dominated by the long-range Coulomb tail.
	
	\subsection{Non-separable systems}
	\label{sec:nSep}
    
    	We now extend the multi-particle WMC method to electron-hole pairs which interact via a reciprocal  soft Coulomb potential \textit{and} with external single-particle potentials, acting separately on the electron and the hole, $V_{e}(x_e(u))$ and $V_{h}(x_h(u))$, respectively.  As discussed in the introduction, in a condensed matter system these could represent electrostatic potentials provided by impurities, defects, alloy fluctuations, metallic gates, etc. 
        
        Keeping with the CM and relative coordinate representation, the Hamiltonian reads
		\begin{equation}
			\Ham = \frac{\widehat{\PP}_{R}}{2M} + \frac{\widehat{\PP}^{2}_{r}}{2\mu} - \frac{\alpha}{\sqrt{\widehat{r}^2 + d^2}} + V_{\text{ext}}(\widehat{r}, \widehat{R})\,.
		\end{equation}
The external potential couples relative and CM coordinates in $V_{\text{ext}}(\hat{r}, \widehat{R}) = V_{e}\big(\widehat{R} + \frac{\widehat{r}}{2}\big) + V_{h}\big(\widehat{R} - \frac{\widehat{r}}{2} \big)$. In other words, external potentials break the translational invariance of the system.  As a result, the full DoFs for the system must be treated, and we would not in general expect a continuous spectrum of energies associated to the free-particle nature of the CM coordinate. From the numerical point of view, exact diagonalisation techniques significantly increase in computational complexity, particularly in higher-dimensional systems, and efficient and accurate numerical methods are desirable. In the following sections, we extend our method to simulate excitons in the presence of two classes of external potentials, and later evaluate its computational efficiency relative to the diagonalisation algorithm.

	\subsubsection{Soft-Coulomb interaction with square potentials}
	\label{sec:softCoulombSqrWells}
		
        We first consider external square potentials along the coordinate of each particle, both in 1D (wires) and 2D (planes). An electrostatic field, of whatever origin, will couple with opposite signs to the two particles of the exciton -- for example, it acts as a well to the electron and as a barrier to the hole. For simplicity, we take both potentials to have width $L$  ($L \times L $) in 1D (respectively, 2D) centred at the origins of the two wires (respectively planes). With this choice of coordinates, the total interaction reads
		\begin{align}
		    \nonumber V(x_{e}(u), x_{h}(u)) &=  V_{\text{SW}}(x_{e}(u)) + V_{\text{SB}}(x_{h}(u)) \\
            &+V_{\text{SC}}(|x_{e}(u)-x_{h}(u)|) 
			\label{eq:Soft-CoulombSqrWells}
		\end{align}
		where (in 2D)
		\begin{equation}
			\displaystyle V_{\text{SW}}(x_{e}(u))={\begin{cases}V_{e}&  |x_e| <\frac{L_x}{2},\, |y_e|<\frac{L_y}{2}\\0&{\text{otherwise}}\end{cases}}
			\label{eq:SqrWell}
		\end{equation}
		and
		\begin{equation}
			\displaystyle V_{\text{SB}}(x_{h}(u))={\begin{cases}V_{h}& |x_h|<\frac{L_x}{2},|y_h|<\frac{L_y}{2}\\0&{\text{otherwise}}\end{cases}} \label{eq:SqrBarrier}
		\end{equation}
(the reduction to 1D is obvious). Here, $V_e<0$ represents the depth of the well and $V_h>0$ the height of the barrier.  With this configuration we expect the possibility of having bound states localised at the barrier/well potentials, since if the electron is trapped by the well, the hole, which has a continuum spectrum, will also be trapped by the reciprocal attraction. Studies of exciton scattering states have indeed confirmed that electron-hole bound states localise at the well / barrier interfaces, which can also be described as an attractive single-particle self-energy potential, due to virtual transitions to excited excitonic states, localised at the well/barrier interfaces \cite{WOS:000403989200009, WOS:000383146200007}.

We use the same WMC representation of the propagator, validated simulating the full DoFs of a two-particle model for separable systems, adapted for the new single-particle potentials, reading
		\begin{align}
			\hspace{-0.5em} &K(x_e^{\prime},x_h^{\prime}, x_e,x_h;T) = K_{0,e}K_{0,h} \nonumber\\
            \hspace{-0.5em}& \times \Big\langle \e^{ -T\int_{0}^{1}du \, \big(V_{\text{SC}}(|x_{e}(u)-x_{h}(u)|) + V_{\text{SW}}(x_{e}(u)) + V_{\text{SB}}(x_{h}(u))\big)} \Big\rangle\,.
			%\hspace{-0.5em}& \times \Big\langle \e^{ -T\int_{0}^{1}du \, \big(\frac{-\alpha}{\sqrt{|x_{e}(u)-x_{h}(u)|^2 +d^2}} + V_{\text{SW}}(x_{e}(u)) + V_{\text{SB}}(x_{h}(u))\big)} \Big\rangle\,.
			\label{eq:WMCpropIXSqrWells}
		\end{align}
		We implement the external potential programmatically by checking when each trajectory lies within its respective well as defined in \eqref{eq:SqrWell} and equation \eqref{eq:SqrBarrier} respectively. Note that in contrast to the diagonalisation (and other algorithms that discretise space(-time), we do \textit{not} suffer a spatial resolution limitation, as finite differences discretisation do -- we discretise the trajectories' parameters, but retain a spatial continuum (our only concern would be ensuring a large enough $N_{p}$ so as to sample the spatial coordinates representatively).  
		
\begin{figure}
		\centering
		\includegraphics[height=0.3\textheight]{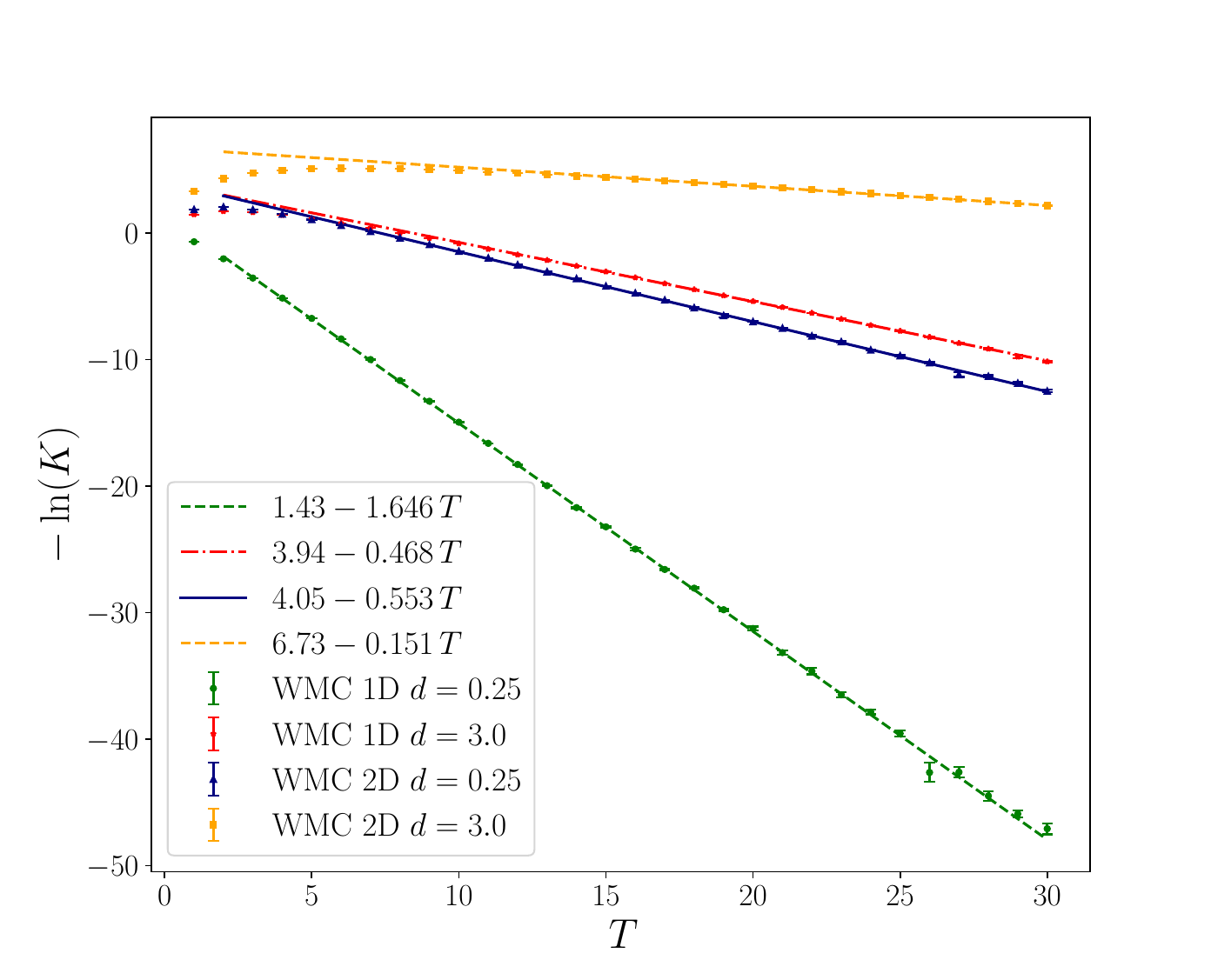}
		\caption{WMC estimates for the logarithm of the kernel for two particles with soft Coulomb interaction and external square potentials vs time $T$, in 1D and 2D, with their respective linear fit. Potential parameters:  $L_{x}=L_{y}=1$, $V_{e}=-1$ and $V_{h}=1$. Parameters of the calculation:  $N_L=5000^2$ , $N_p=20000$.}
		\label{Fig:IXSqrWells}
	\end{figure}

		In Fig.~\ref{Fig:IXSqrWells} we report the WMC estimations of the logarithm of the propagator in 1D and 2D for selected values of $d$ and for $V_e=-V_h$. For 2D systems the non-linear behaviour for small values of $T$ turns into a linear dependence when $T$ is sufficiently large, when the ground state dominates (as we expect a mass gap from the discrete part of the spectrum this late time region should behave as described in(\ref{eq:PropEspectralDecomp})). To extract $E_0$, a time window is determined and a straight line is fitted to the points, $E_0$ being its slope. These linear fits are also shown in Fig.~\ref{Fig:IXSqrWells}. 
	
\begin{figure}
    		\centering
    		\includegraphics[height=0.3\textheight]{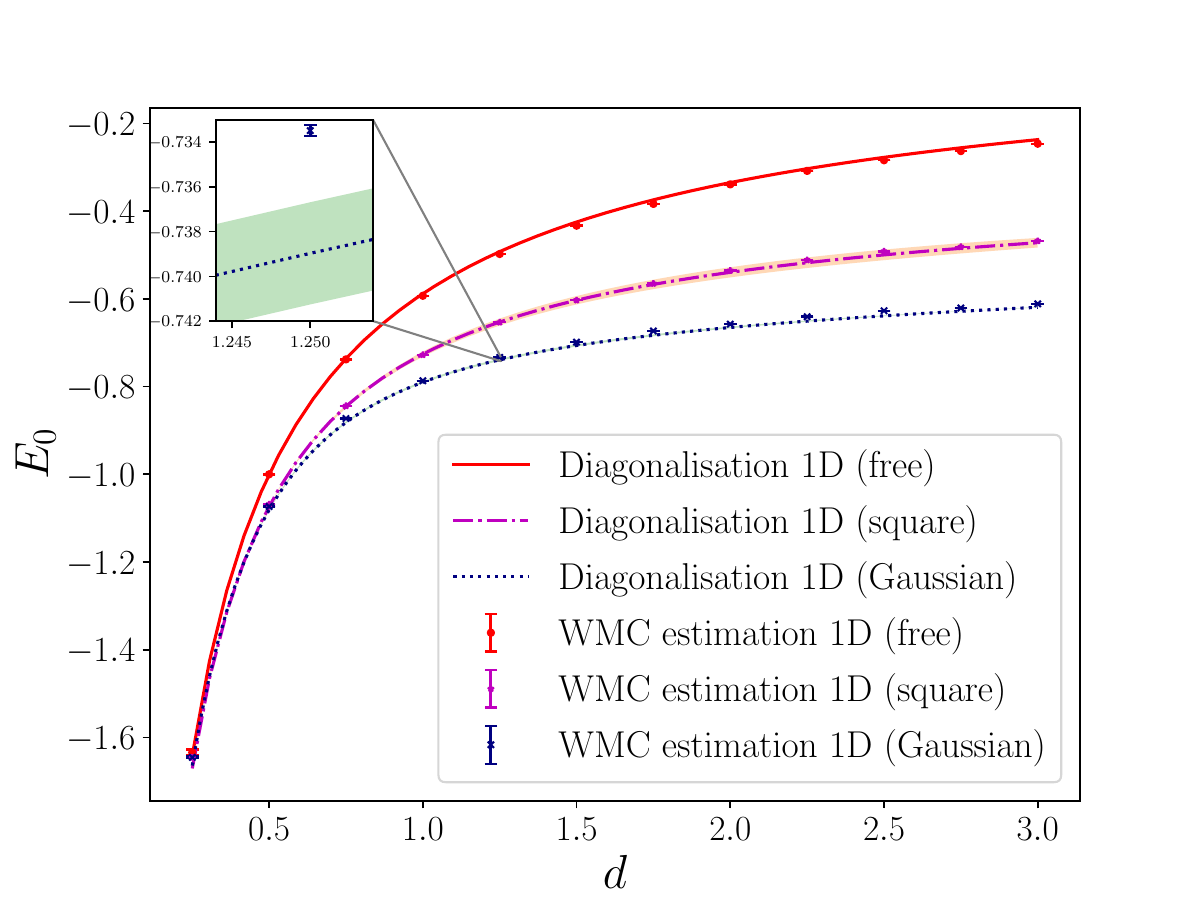}
    		\caption{Ground state energy of two particles in 1D with soft-Coulomb interaction and square or Gaussian external potentials vs cut-off parameter $d$. Potential parameters: $V_e=-1$, $V_h=1$, $L=1$ , $\lambda_e=\lambda_{h}=1$ . Numerical calculation parameters: $N_L=5000^2$, $N_{T}=900$  for square potential; $N_L=4480^2$, $N_{T}=300$  for Gaussian potential.  $N_p=20000$ . Inset: zoom at  $d=1.25$ for the 1D Gaussian potentials case.}
    		\label{Fig:1DIXNonSeparableSys}
        \end{figure}

		In Fig.~\ref{Fig:1DIXNonSeparableSys} we report our estimations of the WMC ground state energy as a function of the cut-off parameter $d$ for free excitons, excitons coupled to square potentials and to Gaussian potentials (to be discussed in Sec.~\ref{sec:softCoulombGaussianPotentials}) in 1D systems. For comparison, we also report there the results obtained by diagonalisation. We observe that the presence of square  potentials tends to decrease the exciton ground state energy compared to the free exciton case for values of $d\ge0.5$. As discussed above, this is expected from exciton localisation at the well boundaries \cite{WOS:000383146200007}. In the limit $d\to0$, the Coulomb interaction dominates and the external potentials have little effect on the ground state energy, so we recover the case of a free exciton -- see Fig.~\ref{Fig:softCou_groundEnergy} and the discussion below. 
        
		The results in Fig.~\ref{Fig:1DIXNonSeparableSys} show that the WMC estimation and numerical diagonalisation agree on the dependence of the ground state energy's on $d$ and its order of magnitude for all three systems. The majority of the results are also quantitatively compatible with one another within their relative error -- however for larger $d$ (especially in the presence of the Gaussian external potentials outlined below), some of the data points indicate predictions that differ by more than $1\sigma$ (see representative inset). In this region the Coulomb interaction is weaker and varies more slowly throughout its domain -- we suggest that a 
faithful sampling of the external potentials becomes more important, and achieving this would require a much larger choice of $N_L$ and $N_{p}$ so as to resolve its long tail.
	
\begin{figure}
		\centering
		\includegraphics[height=0.275\textheight]{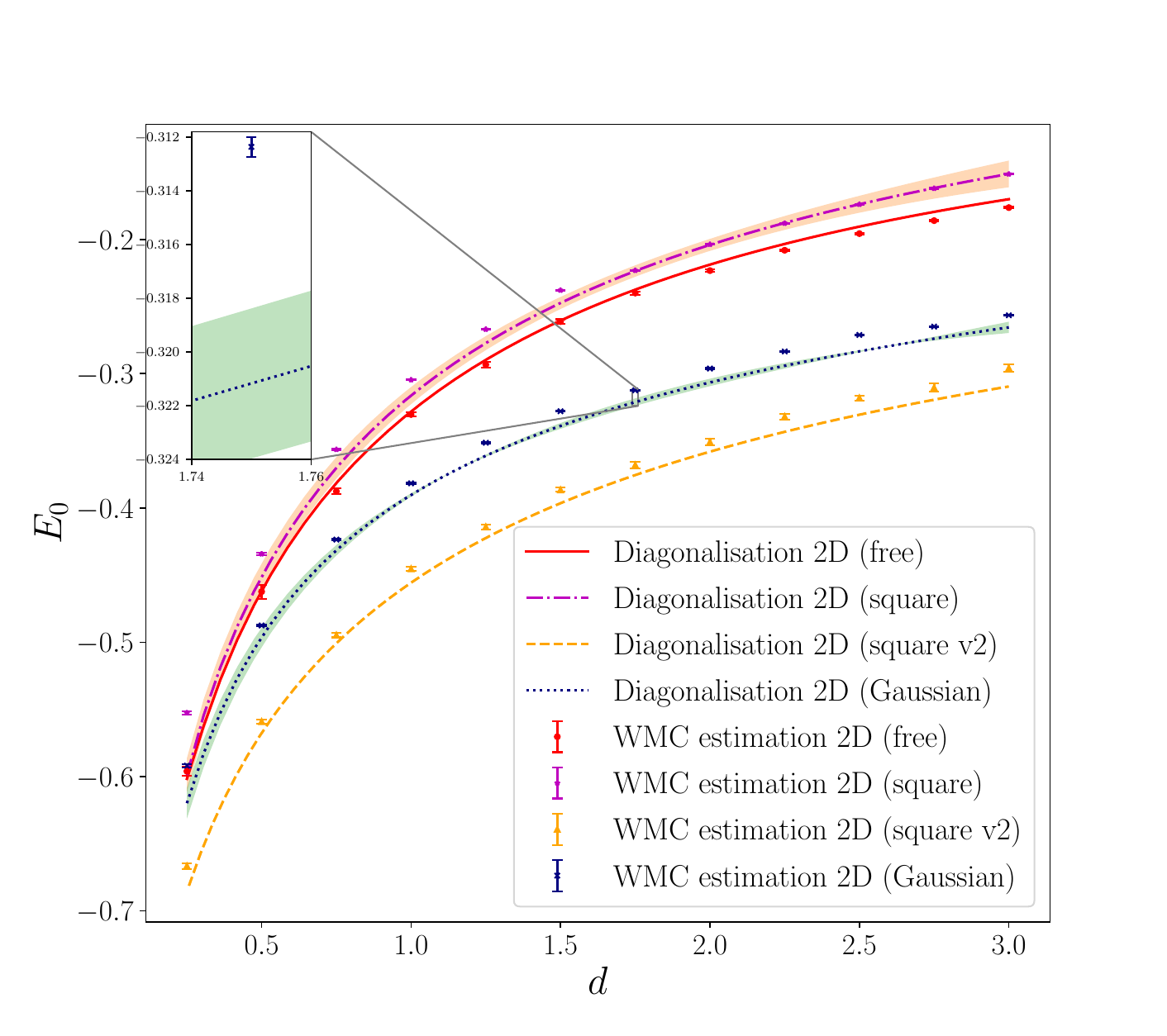}
		\caption{Ground state energy of two particles in 2D with soft-Coulomb interaction and external potentials vs the cut-off parameter $d$. For WMC: $N_L=5000^2$ (square potential) and $N_L=4480^2$; $N_p=20000$ for Gaussian potentials. Potential parameters:   $L_{x}=L_{y}=1$, $\lambda_e=\lambda_{h}=1$, $V_e=-1$, $V_h=1$, and $V_e=-2$, $V_h=2$ (labelled v2). For diagonalisation:  $N_{T}=60^{4}$ (square potential),  $N_{T}=80^{4}$ (Gaussian potentials), $N_{T}=84^{4}$ (Gaussian potentials, v2) and. Inset: zoom to the region at $d=1.75$ for the 2D Gaussian potentials case.}
		\label{Fig:2DIXNonSeparableSys}
	\end{figure}

		In Fig.~\ref{Fig:2DIXNonSeparableSys} we show a similar plot as Fig.~\ref{Fig:1DIXNonSeparableSys}, now for 2D systems. Again the dependencies and orders of magnitude of the WMC numerical results are similar to exact diagonalisation, especially for smaller values of $d$. The discrepancy is more marked with respect to the 1D case, however, especially for square potentials. 
		%-- the inset shows a greater difference between the numerical results, despite larger error bars.
		In particular, examining the convergence plots for the diagonalisation method in Appendix \ref{sec:ApTimeScaling}, we find that the estimates of the ground state energy do not converge monotonically as we increase the number of grid points $N$ used. This is to be expected, perhaps, since the diagonalisation method (which, we recall, works entirely in position space) with non-smooth potentials can develop sensitivity to the discretisation parameter(s) ($N_{x}$, $N_{y}$), as the method struggles to resolve the edges of the potential wells  as the grid points jump from one side of the edge to the other. 
        
        Moreover we notice from Fig.~\ref{Fig:2DIXNonSeparableSys} that application of external square potentials  with $V_{e}=-1$ and $V_{h}=1$ increases the ground state energy with respect to the free exciton case. If, however, we  make the attractive well deeper and the barrier higher, e.g.~$V_{e} = -2$ and $V_{h} = 2$ -- labelled as V2 in Fig.~\ref{Fig:2DIXNonSeparableSys},  the ground state energy is lower than in the free exciton case.  This behaviour can be understood as the competing effects of the potential well acting to localise the electron, while the barrier scatters the hole away from it -- which tends to increase the Coulomb energy and quantise the (positive) CM energy once translational invariance is removed -- with the reduction in energy associated to the trapping of the electron in that well. As can be seen, the latter dominates for deeper ($V_{e}$, $V_{h}$) (or wider, $L$) wells and weaker Coulomb coupling, $\alpha$. We verified this with a variational approach to estimate the ground state energy: the optimal wavefunctions narrowly confine the electron to the potential well but balance reducing the hole's overlap with the barrier against an overall increase in inter-particle distance. In 1D the Coulomb interaction still dominates the determination of the system's as $d \to 0$, as the Hamiltonian's spectrum is unbounded from below in this limit; hence the ground state energies converge on that of the purely Coulombic system for small $d$. This is not the case in 2D where the purely Coulombic system has a finite binding energy so that the external potentials can influence the ground state energy of the combined system.

	\subsubsection{Soft-Coulomb interaction with Gaussian potentials}
	\label{sec:softCoulombGaussianPotentials}
	
    Now we consider excitons in the presence of Gaussian external  potentials. Smooth potential wells and barriers are clearly more realistic models for material systems, e.g.~ in experimental studies of tunnelling processes. We consider the potential 
    \begin{align}
        \nonumber V(x_{e}(u), x_{h}(u)) &=  V_{e}\,\e^{-\lambda_{e}|x_{e}(u)|^{2}} + V_{h}\,\e^{-\lambda_{h}|x_{h}(u)|^{2}} \\
            &+V_{\text{SC}}(|x_{e}(u)-x_{h}(u)|) \,,
    \end{align}
    where $V_e<0$ and $V_h>0$ are the Gaussian well depth and barrier height, respectively, and $\lambda_{e}$, $\lambda_{h}$ the (inverse) Gaussian widths. This system is, of course, non-separable and no closed form solution exists for its spectrum or eigenstates. The WMC representation of the propagator for the system with this interaction reads
    \begin{align}
        \nonumber &K(x_e^{\prime},x_h^{\prime}, x_e,x_h;T) = K_{0,e}K_{0,h}\\
        &\Big\langle \e^{ -T\int_{0}^{1}du \, \big(V_{\text{SC}}(|x_{e}(u)-x_{h}(u)|) +V_{e}\,\e^{-\lambda_{e}|x_{e}(u)|^{2}} + V_{h}\,\e^{-\lambda_{h}|x_{h}(u)|^{2}}\big)} \Big\rangle\,,
        \label{eq:WMCpropIXGaussPot}
    \end{align}
    which presents no obstacle to simulation using the numerical algorithms presented above. Despite changing the shape of the external potentials, we  expect that results obtained in the previous subsection are qualitatively recovered using one attractive and one repulsive Gaussian potential -- we are interested, then, in the effect of the greater spatial extent of these potentials. 
    
    From Figures \ref{Fig:1DIXNonSeparableSys} and \ref{Fig:2DIXNonSeparableSys} we note that the Gaussian potentials confine the particles more effectively than square potentials for comparable parameters; we interpret this as the softer decay of the tails of the Gaussian wells, increasing their spatial extent compared to the sharp edges of the square well, thereby allowing a more favourable reduction in energy. In the regime when $d<0.5$, we observe that the soft Coulomb interaction dominates again and the estimations of the ground state energy with Gaussian and square potentials are very similar. In 1D (Fig.~\ref{Fig:1DIXNonSeparableSys}), there is good agreement between both numerical approaches for lower values of $d$ -- as can be seen in the inset, however, they are not quite consistent within the estimated errors for $d \gtrsim 1$. Then, for the 2D case, Fig.~\ref{Fig:2DIXNonSeparableSys} shows a greater misalignment  between the WMC estimate and the diagonalisation result with a grid of $N_{T}=80^{4}$ points, despite both methods having small errors. 
    
    Here we argue in favour of the WMC estimation by examining the limit $d \to 0$. In conducting diagnostic tests of the stability of our results we found that there is considerable variation in the estimations of the ground state energy as we vary the grid size used in the diagonalisation method (see also Appendix \ref{sec:ApTimeScaling}), which suggests that the results are not robust (or at least, it is unclear that we are approaching the continuum limit).  It is possible, then, that application of the diagonalisation method with significantly more than $N_{T}=80^{4}$ total grid points would be necessary to estimate the ground energy and improve the agreement between the two methods. However, doing this would come at a much higher cost of computing resources and time (see discussion in Appendix \ref{sec:ApTimeScaling}) and would require recalibration of the grid used to obtain the good estimations of energies for larger values of $d$. Instead we limit ourselves to presenting the evidence in favour of the WMC simulation here, which is seen to produce consistent predictions for the ground state energy that correctly reduce to their expected limiting value, and that are stable with respect to further refinement of the parameters, $N_{L}$ and $N_{P}$ used in the simulations.  

\subsubsection{Numerical efficiency and robustness}
	
    We conclude this section by discussing questions of computational complexity (with reference to the results presented in Appendix \ref{sec:ApTimeScaling}). The WMC algorithm simulation time essentially scales linearly with the number of loops, $\tilde{N}_{L}$. We compare this to the simulation time of the diagonalisation code, which (a) scales worse (almost quadratically) with number of discretisation points in $D > 1$ dimensions and (b) automatically requires the total number of discretised spatial points to increase with dimensionality, with the effect of increasing the scaling exponent with respect to the number of points used. In both cases, as the dimensionality is increased, maintaining a comparable spatial resolution requires the number of trajectories, $\tilde{N}_{L}$, (respectively discretisation grid density, proportional to $1/N$), to be increased, but with the favourable scaling of the WMC simulation this can be achieved more efficiently for large values of these parameters. In this sense we argue that the WMC technique avoids the familiar problems of scaling numerical solvers of the Schrödinger equation to higher numbers of spatial dimension. We return to this question for the extension to 3-dimensional systems at the end of the proceeding section.
	    
	As a final consideration, note that above we have chosen attractive and repulsive potentials, either square or Gaussian, with identical strengths $V_e = - V_h$. Of course, this is not a limitation of the method by all means, and systematic investigation with $|V_e|\ne|V_h|$ would be possible. The same applies to masses, which are taken as identical in this work. However, note that for identical masses and equal but opposite potentials, electron and hole wavefunctions are identical. Therefore, the average external potential felt by the pair, which is the lowest order external potential contribution to the exciton dynamics, vanishes \cite{WOS:000383146200007}. In other words, with this choice of parameters, localisation  is only due to the correlation effects induced by the reciprocal (here, soft Coulomb) interaction. Hence, this condition is the more stringent testbed for the WMC method. 
	
	\subsection{3D quantum confined exciton model}
	\label{sec:ExcZdir}

    Realistic systems in condensed matter with 1D or 2D dynamics necessarily have a finite extension in confined directions. While strictly, idealised 1D or 2D systems are still relevant  when low-energy phenomena are considered (e.g.~in many-body physics of the electron gas) and only the ground state in the confined direction is relevant, it is often necessary to consider the full 3D nature of the system for quantitative predictions, e.g.~of inter- or intra-band optical transitions. 
        
        Of course, dealing with strictly 1D or 2D systems complies with computational limitations which generally  scale unfavourably with dimensionality. However, one of the advantages of the WMC formalism (and Monte Carlo integration techniques in general) is that we can increase the dimensionality of the system with relative ease (compare, for example, the  soft scaling behaviour with respect to dimension exhibited for WMC in Appendix \ref{sec:ApTimeScaling}) with the (almost) quadratically grown in computational cost of the diagonalisation code. 
        
	 Therefore, as a non-trivial application in 3D, we study here the dimensional reduction of this system outlined in \ref{sec:IndirectExcitons} as a limiting process of strong and narrow external potential.  Such a system would require large resources of physical memory and computing time (see Appendix \ref{sec:ApTimeScaling}) using diagonalisation techniques. 
		%computational time scales with grid discretisation to the sixth power  
		This makes the continuum limit difficult to approach to effectively resolve the confining potential required for the dimensional reduction.
		
		We consider a 3D model of an electron-hole pair with the soft Coulomb interaction, with coordinates $x_e, x_h$ in the free $x$ direction, and with confining potentials in the $z$ direction consisting of two square wells which are separated by a distance $d$ (\cite{bastard1982exciton} for a variational study in  GaAs-GaAlAs systems). Such systems are routinely obtained in semiconductor physics by proper modulation of materials in a heterostructure. Here,  the parameter $d$ takes the meaning of the distance between the median point of the two wells. Each well has a finite extension $L_z$  and a well depths $V_e$, $V_h$. We shall study the ground state as a function of these parameters below.

                The total intentional term for the two-particle Hamiltonian reads

		\begin{align}
		    \nonumber V(x_{e}(u), x_{h}(u)) &=  V_{\text{SW}}\big(z_e(u)+\frac{d}{2}\big) + V_{\text{SW}}\big(z_h(u)-\frac{d}{2}\big) \\ &-\frac{\alpha}{|x_{e}(u)-x_{h}(u)|} 
		\end{align}
		where $V_{\text{SW}}$ is defined as in \eqref{eq:SqrWell}, for the z direction, with width $L_z$ taken to be equal for both quantum wells and $V_{e,h} < 0$. The electron and the hole, therefore, both feel a confining potential, and can be confined in different layers (which, in a real system,can be easily achieved by breaking the symmetry along $z$  by a small electric field, not included in $V$ for simplicity) and the soft Coulomb interaction parameter $d$ takes the meaning of the mean separation between the two particles. 
        
        The WMC propagator reads
		\begin{align}
			\nonumber K&(x_e^{\prime},x_h^{\prime}, x_e,x_h;T)=K_{0,e}K_{0,h} \\
			&\times \Big\langle \e^{ -T\int_{0}^{1}du \, \big(\frac{-\alpha}{|x_{e}(u)-x_{h}(u)|} + V_{\text{SW}}(z_e(u) + \frac{d}{2}) + V_{\text{SB}}(z_h(u) - \frac{d}{2})\big)} \Big\rangle\,.
			\label{eq:WMCprop3DIXSqrWells}
		\end{align}
		where the boundary conditions in the third direction are $z_{i}(0) = \mp \frac{d}{2} = z_{i}'(0)$ for $i = e, h$. We expect to recover the 2D soft Coulomb case for the limit when the widths of the square potentials are small and their strength very large, i.e., when $L_z\to0$ and $V_{e,h}\to-\infty$. Throughout this section we fix $d = 2$.
		
\begin{figure}
			\centering
			\includegraphics[width=0.5\textwidth]{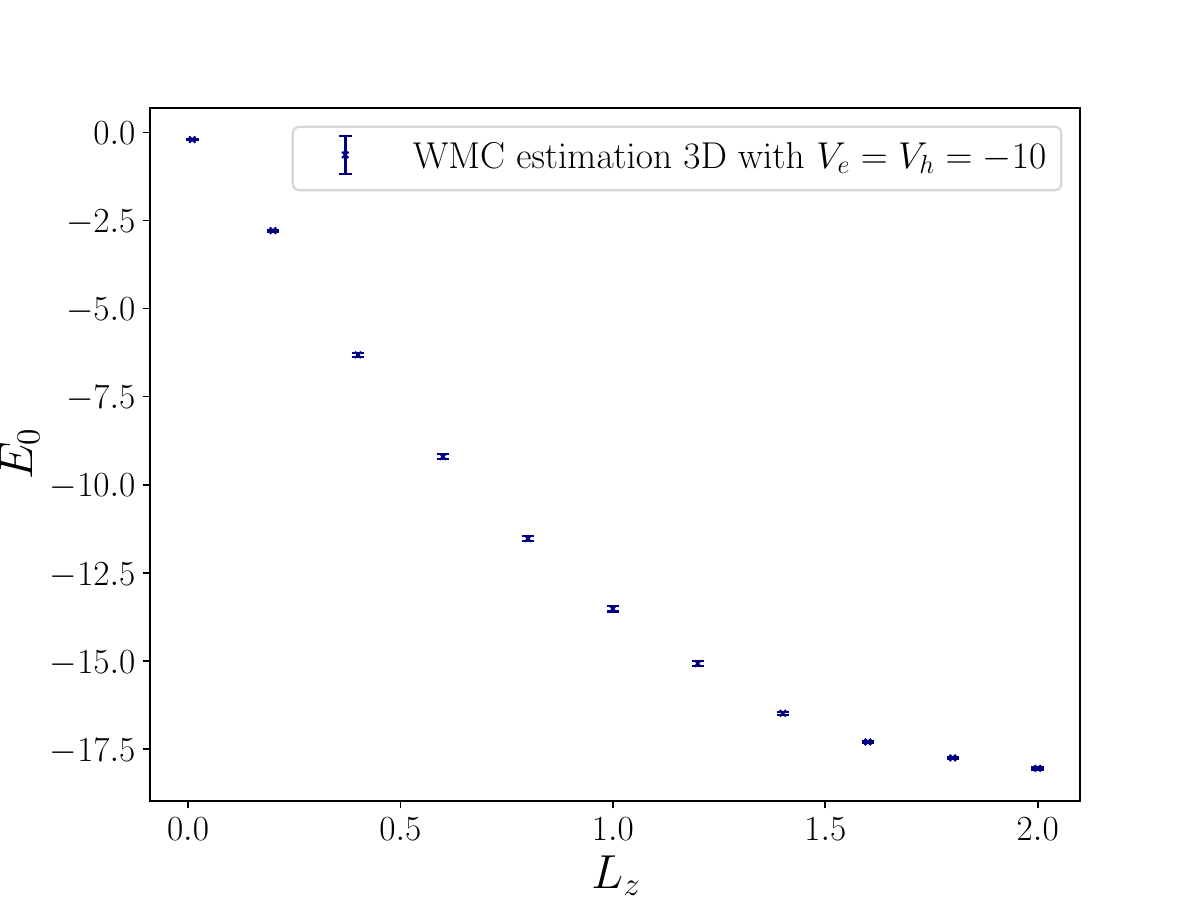}
			\caption{Ground state energy of two particles  with soft Coulomb interaction and square wells confinement in the z direction.  We used $N_L=7000^2$ loops, $N_p=10000$ points per loop, $m_{e}=m_{h}=1$, $\alpha=1$, $V_e=-10$, $V_h=-10$ and $d=2$ .}
			\label{Fig:3DsoftCouSqrWells_Lz}
		\end{figure}

		 In Fig.~\ref{Fig:3DsoftCouSqrWells_Lz} we plot the exciton ground state energy as a function of the width $L_z$ for a value of $V_{e,h} = -10$ that we found to be sufficiently ``large'' in our units. To explore the aforementioned limit, we took the well width down to $L_z=0.01$, resulting in an estimate of $E_{0}=-0.204(2) \pm 0.001(7)$, a value close to the estimate for a purely two dimensional system without external potentials, $E_{0}=-0.2232(3) \pm 0.0008(9)$. Away from this limit, we observe the effect of the extent of the square wells' partial confinement in the monotonically decreasing values of the binding energy as $L_z$ increases\footnote{Running simulation of the diagonalisation code with $N_{T}=20^6$ points on the grid, we could only corroborate the scale in the $E_0$ axis values.}, in agreement with the qualitative behaviour observed in \cite{bastard1982exciton}. 
		
        %\begin{figure}
		%	\centering
		%	\includegraphics[width=0.5\textwidth]{3DsoftCoulombSqrWells.pdf}
		%	\caption{Ground state energy of 2 particles  with soft Coulomb interaction and square wells confinement in the z direction.  We used $N_L=7000^2$ loops, $N_p=10000$ points per loop, $m_{e}=m_{h}=1$, $\alpha=1$, $V_e=-10$, $V_h=-10$ and $d=2$ .}
		%	\label{Fig:3DsoftCouSqrWells_Lz}
		%\end{figure}

        %\begin{figure}
        %    \centering
        %    
        %    \subfloat[Ground state energy vs. $L_z$ with $V_e=V_h=-10$.\label{Fig:3DsoftCouSqrWells_Lz}]{
        %        \includegraphics[width=\columnwidth]{3DsoftCoulombSqrWells_Lz.pdf}
        %    }
        %    
        %    \vspace{0.3cm} % Adjust vertical spacing
        %    
        %    \subfloat[Ground state energy vs. $V_e=V_h$ with $L_z=1$.\label{Fig:3DsoftCouSqrWells_VeVh}]{
        %        \includegraphics[width=\columnwidth]{3DsoftCoulombSqrWells_VeVh.pdf}
        %    }
        %    
        %    \caption{Ground state energy of 2 particles in 3D with Coulomb interaction and square wells confinement in the $z$ direction. Parameters: $N_L=7000^2$ loops, $N_p=10000$ points per loop, $m_e=m_h=1$, $\alpha=1$, $d=2$.}
        %    \label{Fig:combined_plots}
        %\end{figure}
		
\begin{figure}
			\centering
			\includegraphics[width=0.5\textwidth]{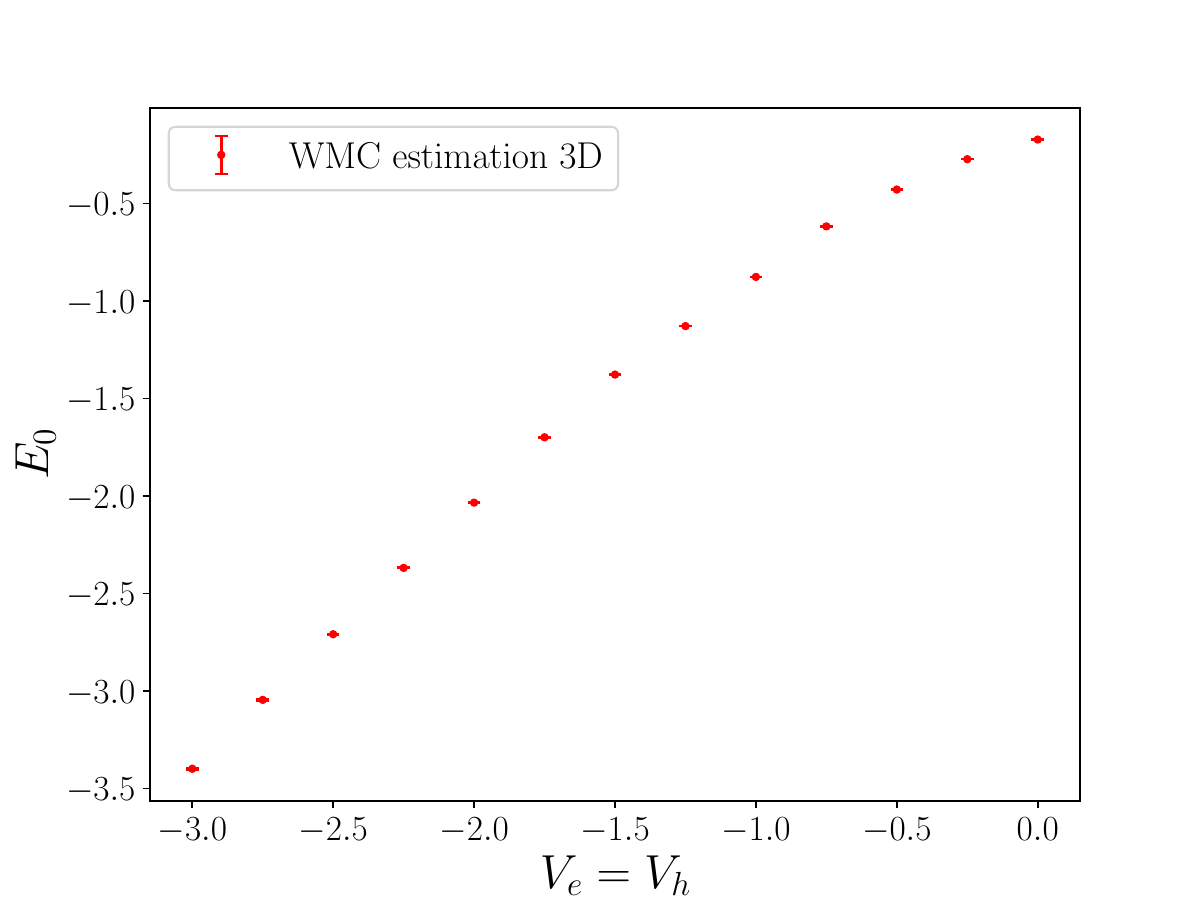}
			\caption{Ground state energy comparison of two particles in 3D with a Coulomb interaction and square wells confinement in the z direction. We used $N_L=7000^2$ loops, $N_p=10000$ points per loop, $m_{e}=m_{h}=1$, $\alpha=1$, $L_z=1$ and $d=2$ for each different value of the deep of the square wells $V_e=V_h$.}
			\label{Fig:3DsoftCouSqrWells_VeVh}
		\end{figure}

		In Fig.~\ref{Fig:3DsoftCouSqrWells_VeVh} we plot the ground state energy for different values of the depth (strength) of the square wells $V_{e}=V_{h}$ at fixed $L_{z} = 1$. Here we also observe the monotonically decreasing behaviour of the ground state energy as the depth increases, with a limiting value of $E_{0}=-0.1725(4) \pm 0.0003(2)$ for $V_{e}=V_{h}=0$. This, of course corresponds to an  exciton in 3D  (``excitons in the bulk'') with  $d=2$. 
        
        These results show that the WMC algorithms for the two-particle system extend well to 3D quantum systems and that the ground state energy can be estimated without a significant increase in numerical uncertainty. As shown in Appendix \ref{sec:ApTimeScaling}, the simulations also scale well with respect to $\tilde{N}_{L}$,  with simulation time growing similarly to the 2D and 1D WMC simulations, and more favourably than the diagonalisation method.
		
	\subsection{Three particle simulations}
	\label{sec:Trions}
	
		Although the theoretical development of the WMC algorithms in section \ref{sec:Theory} considered an arbitrary number of particles, we have up until now focused on simulating two-particle systems (excitons) subjected to different interactions. Of course there are various competing numerical algorithms for finding the spectrum of energies of the quantum systems, straightforward diagonalisation being just the most simple example (albeit we have argued they do not scale well as compared to the present WMC approach).  However, it is also increasingly challenging to simulate systems with a greater number of particles, for which, of course, various approximate approaches have been developed to handle the increased complexity of a greater number of DoFs.

		In this subsection, we consider the case of a three-particle system modelling a ``trion'' (also known as a charged exciton): a bound state consisting of two electrons and a hole (negative trion $X^{-}$) or two holes and one electron (positive trion $X^{+}$). This is a natural extension to studies of excitons presented above and provides a proof of the principle that the WMC approach developed here extends naturally to higher numbers of particles without spoiling the gentler scaling with respect to number of trajectories generated. 
        
        The existence of trion excitations was theoretically predicted in semiconductors by Lampert \cite{Lampert:1958} and, a few decades later, first observed experimentally in semiconductor quantum wells \cite{Kheng:1993} -- more recently, trionic quasi-particle excitation have also been observed in carbon nanotubes \cite{Matsunaga:2011,JinSungPark:2012}. Due to the complexity of analytically treating the three body problem, theoretical efforts to describe trions analytically have largely been conducted using approximate methods such as the variational principle; this method has been used to study positive and negative trions in semiconductor quantum wells \cite{Stebe:1997,Stebe:2000,Riva:2001,Sergeev:2001}, carbon nanotubes \cite{Ronnow:2009, Ronnow:2010,marchenko:2012stability}, and quantum wires \cite{Szafran:2005}.  
		
		To test the capacity of the WMC formalism  here we consider three particles with inter-particle interactions only (free trion) using two models proposed in the literature, both being effectively 1D systems. We emphasise that our model here treats identical particles (holes or electrons) as \textit{distinguishable} -- a more refined simulation would consider the required antisymmetry in these particles' wavefunctions.
		
		 In the first model, taken from \cite{marchenko:2012stability}, the stability for positive trions ($X^{+}$) in one-dimensional zigzag carbon nanotubes was studied. There the trion interaction considered reads  (we continue with the shorthand notation $x_{e}$ for the electron position and $x_{h_1}$, $x_{h_2}$ for the positions of the holes)
		\begin{align}
			\nonumber V &= V_{\text{SC}}(|x_{e}(u)-x_{h_1}(u)|) + V_{\text{SC}}(|x_{e}(u)-x_{h_2}(u)|) \\
			& - V_{\text{SC}}(|x_{h_1}(u)-x_{h_2}(u)|)\,,
			\label{eq:TrionSoft-Coulomb}
		\end{align}
		where $V_{\text{SC}}(|x_i(u)-x_j(u))|$ is defined in equation \eqref{eq:Soft-Coulomb} for 1D, but we should reinterpret the parameter $d$ as now parameterising the nanotube diameter. Hence for this system the WMC estimation of the propagator reads
		\begin{align}
			\hspace{-3mm} \nonumber &K(x_{e}^{\prime},x_{h_1}^{\prime}, x_{h_2}^{\prime},x_{e},x_{h_1}, x_{h_2};T) = K_{0,e}K_{0,h_1}K_{0,h_2}\\
            \hspace{-3mm} &\times\Big\langle  \e^{-T\int_{0}^{1}du \, \big[V_{\text{SC}}(|x_{e}(u)-x_{h_1}(u)|) + V_{\text{SC}}(|x_{e}(u)-x_{h_2}(u)|) \big]} \nonumber\\
			& \hspace{4em}\times\e^{T\int_{0}^{1}du\,V_{\text{SC}}(|x_{h_1}(u)-x_{h_2}(u)|) } \Big\rangle\,.
            %\hspace{-3mm} &\times\Big\langle  \e^{-T\alpha\int_{0}^{1}du \, \big[-\frac{1}{\sqrt{|x_{e}(u)-x_{h1}(u)|^2 +d^2}} - \frac{1}{\sqrt{|x_{e}(u)-x_{h2}(u)|^2 +d^2}}\big]} \nonumber\\
			%& \hspace{4em}\times\e^{-T\alpha\int_{0}^{1}du  \frac{1}{\sqrt{|x_{h1}(u)-x_{h2}(u)|^2 +d^2}} } \Big\rangle\,.
			\label{eq:WMCpropSoftCoulombTrion}
		\end{align}
		We also consider negative trions ($X^{-}$) in quantum wires, inspired by \cite{Szafran:2005},  where the DoFs perpendicular to the wire are under the influence of a harmonic oscillator-type confinement potential of lengths $l_{e_1}$, $l_{e_2}$ and $l_{h_2}$. If the confinement is assumed to be strong, the lateral DoFs can be integrated, giving way to an effective interaction in 1D that models the trion's motion on the quantum wire. Specifically, integrating the Coulomb interaction along the lateral direction assuming the particles are in the \textit{ground state} of the harmonic trap leads to a series of effective potential in the remaining directions,
		\begin{align}
			\nonumber V &= V_{\text{eff}}(d;\, |x_{e_1}(u)-x_{e_2}(u)|) - V_{\text{eff}}(d;\,|x_{e_1}(u)-x_{h}(u)|) \\
			 & - V_{\text{eff}}(d;\,|x_{e_2}(u)-x_{h}(u)|)\,,
			\label{eq:EffectiveTrion}
		\end{align}
		where we have considered equal electron and hole lateral confinement ($l_{e_i}=l_{h}=d$) and 
		\begin{equation}
			V_{\text{eff}}(d;\, |x_i-x_j|)=\frac{1}{d}\sqrt{\frac{\pi}{2}} \operatorname{erfc}\Big(\frac{|x_i-x_j|}{d\sqrt{2}}\Big) \e^{\frac{|x_i-x_j|^2}{2d^2}}\,.
			\label{eq:EffectiveTrionPotentialForm}
		\end{equation}
		Now the parameter $d$ doubles over as the confinement distance and characteristic fall-off of the effective potential -- this potential offers an alternative to the soft Coulomb potential which is also finite when $|x_i-x_j|=0$, with a much faster (exponential) decay but that is stronger at short distances. For this second system the WMC estimate for the propagator is
		\begin{align}
			\nonumber &K(x_{e}^{\prime},x_{h1}^{\prime}, x_{h2}^{\prime},x_{e_1},x_{e_2}, x_{h};T) = K_{0,e_1}K_{0,e_2}K_{0,h}\\
			\nonumber &\times \Big\langle \e^{ -T\int_{0}^{1}du \, \big[  - V_{\text{eff}}(d;|x_{e_1}(u)-x_{h}(u)|) - V_{\text{eff}}(d;|x_{e_2}(u)-x_{h}(u)|)\big] } \\
			&\hspace{3em}\times \e^{-T\int_{0}^{1}du \, V_{\text{eff}}(d;|x_{e_1}(u)-x_{e_2}(u)|)}\Big\rangle\,.
			\label{eq:WMCpropEffectiveTrion}
		\end{align}
		In both models we have neglected the difference between the effective masses of the electrons and holes in experimental systems so as to simplify the models, so the ratio $\sigma=m_e/m_h=1$. 
		
\begin{figure}
			\centering
			\includegraphics[width=0.5\textwidth]{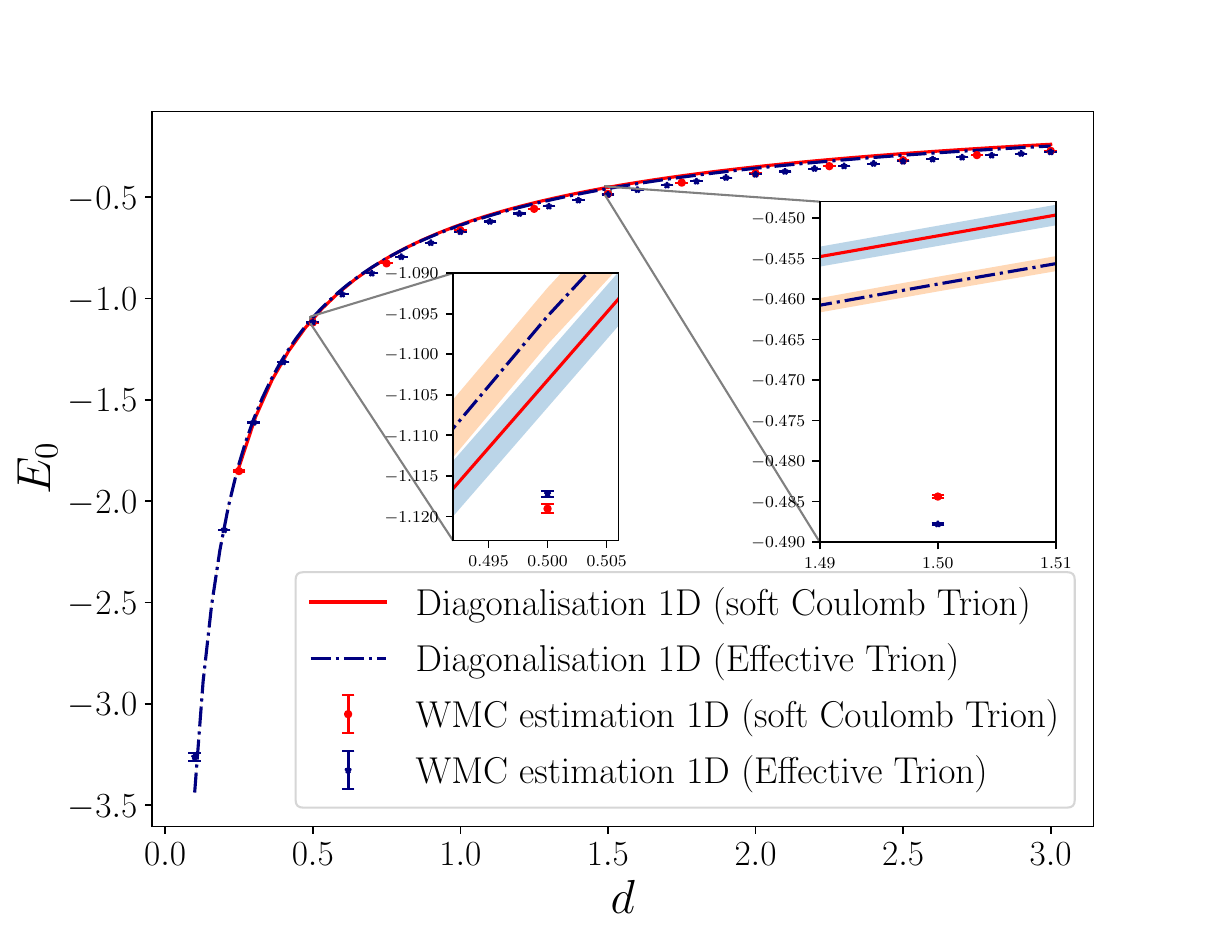}
			\caption{Ground state energy of trion vs $d$ in 1D with soft-Coulomb type potential  and an effective interaction. We used $N_L=5000^3$  and $N_L=10000^3$ respectively, $N_p=5000$, $m_{1}=m_{2}=m_{3}=1$ and $\alpha=1$,  $N_{T}=260^3$ . Insets: zoom to the regions near values $d=0.5\;\&\;d=1.5$ for both potentials.}
			\label{Fig:Trions}
		\end{figure}

		In Fig.~\ref{Fig:Trions} we plot the ground state energy of both models against the parameter $d$ (which has different physical meaning for each system) along with the results provided by a direct numerical diagonalisation of the Schr\"odinger equation.  
		We observe that the two models are in  very good agreement both in WMC estimates and direct diagonalisation result. However, the  diagonalisation method systematically gives ground state energies which are slightly higher then  the WMC results. The discrepancy between the methods also increases with $d$ (see the insets in Fig.~\ref{Fig:Trions}).
			
		Our results with are consistent with Fig.~5 in Ref.~\cite{Szafran:2005} . In Ref.~\cite{Szafran:2005} the ground state energy as a function of the lateral confinement length diverges towards minus infinity  approaching  $d=0$, as a consequence of the singularity of the Coulomb interaction in 1D \cite{RLoudon1959Hydrogen1D,RLoudon2016Hydrogen1D}. The ground state energies have a dependence $d^{\alpha}$, with $\alpha=-0.74$ calculated with the WMC formalism and $\alpha=-0.79$ with the diagonalisation method; these results are close to the exponent calculated in \cite{Szafran:2005}, with $\alpha=-0.83$, using a reduced two particle system for the trions.  Note, however, that we will simulate the constituents of the trion as independent particle trajectories.
        
		In the case of trions in carbon nanotubes, it has been reported in \cite{Ronnow:2009, Matsunaga:2011, marchenko:2012stability} that there is an approximate dependence of the trion's ground state energy on the nanotube diameter, which goes as $E_{X^{\pm}} \approx \beta / d$, where $\beta$ is a positive constant. We used our data to determine whether the dependence on $d$ is consistent with an exponent close to minus one, obtaining in our case a scaling $E_{X^{\pm}} \sim d^{\alpha}$ with an exponent that is approximately $-0.77$ from the WMC data and $-0.83$ from the diagonalisation results. Let us note that these values are in close agreement with those calculated for the case of the soft Coulomb potential. 
		
	It is important to note that, since want to put the WMC calculations to the test, our calculations here are conducted taking into full account the coordinates of the three particles, i.e., we did not separate the CM coordinate, which would of course be possible and convenient in the free case, and is invariably done in literature. In terms of their computational complexity, though, the three-particle system simulated here did not present any obstacles in the WMC simulation due to the increased number of DoFs. Of course, the total number of loops used, $\tilde{N}_{L}$, increases due to the inclusion of the third particle (electron or hole) but as shown in Appendix \ref{sec:ApTimeScaling}, the WMC simulations scale approximately linearly with this parameter, at difference with, e.g.,  the  scaling with the diagonalisation method. As such we favour the WMC simulation of this system, being both easier to scale to higher dimension and in approaching the continuum limit.
	
	%%%%%%%%%%%%%%%%%%%%%%%%%%%%%%%%%%%%%%%%%%%%%%%%%%%%%%%%%%%%%%%%%%%%%%%%%%%%%%%%%%%%%%%%%%%%%
	\section{Conclusions}
	\label{sec:Concl}
    We have extended, in the natural way, Worldline Monte Carlo simulations of quantum mechanical systems to diverse multi-particle cases.  This required generalising the numerical algorithms producing particle trajectories to incorporate additional particle worldlines and to couple these trajectories by both inter-particle interactions and to external potentials. We have done this for local interactions (although the extension to non-local interactions is straightforward) and a range of external potentials. Along the way, we have presented updated algorithms that produce multiple, independent and identically distributed Brownian motion trajectories, generalised a smoothing procedure (Appendix \ref{sec:ApSmooth}) that neatly handles singularities in interaction potentials and carried out a systematic analysis of the computational complexity of the WMC algorithm and its scaling with the number of worldlines produced.
    
    The results of our numerical simulation start with an estimation of the quantum mechanical propagator for the multi-particle system, describing the probability amplitude of propagation between fixed spatial points in a (global) given time. From the late-time estimation of the propagator, however, we can extract the ground state energy of the system (we dealt with an additional complication for separable systems without a mass gap) -- and this could easily be extended to estimating the energy of the first excited state for potentials that retain a parity symmetry as in \cite{UsMonteCarlo}. In this way, we have explored the dependence of the binding energy of exciton-like systems and the ground state energy for these same systems exposed to external potentials, on the cut-off parameter, and we systematically compared our results to analytic calculations, where they exist, and a basic numerical solver based on direct diagonalisation of the system Hamiltonian represented with finite differences in real-space. 
    
We were able to reproduce the analytic results for a simple toy model involving a harmonic inter-particle interaction and accurately simulate models of free excitons and those exposed to external potential wells. We also simulated two systems composed of trions, less commonly studied in the literature, again finding qualitative and quantitative agreement with past results. For all cases, WMC estimates agree well with the benchmarks. For many values of the parameters explored, results coincide within the estimated numerical errors, evaluated as explained in the main text. For some systems there are small discrepancies between the predictions which may indicate that we are underestimating the errors by a factor of order unity. Overall, the demonstrated accuracy of the WMC estimates for a wide range of $d$, which corresponds to the exciton wavefunction sampling very different regions of the soft Coulomb potential, speaks in favour of our method over a variety of different physical situations, even in the particularly challenging case of long-ranged interactions.
    
We have also demonstrated that the WMC approach scales more favourably than the diagonalisation code, both with respect to the discretisation parameter and dimensionality. This is largely due to the fact that, in the latter method, a discretisation in configuration space (or, equivalently, in momentum space) inevitably requires an increase in the number of lattice (equivalently, wavevector) points; instead, the WMC method does not  necessitate this because the worldlines naturally explore all dimensions, due to their Brownian motion nature, and are not constrained to spatial resolutions set by a lattice spacing.  Instead their spatial resolution is related to the number of points per loop, $N_{P}$, and their length, which grows with propagation time $T$. This allowed us to apply the WMC formalism in one-, two- and three-dimensional space, including external potentials, with the 3D simulations prohibitively computationally expensive to achieve -- with our hardware -- for the diagonalisation method. 
    
    One point worth emphasising is that the total number of loops, $\tilde{N}_{L}$, required for accurate numerical estimations is much larger for the multi-particle simulations as for a one-body problem (typically we used $\tilde{N}_{L} = N_{L}^{n}$ trajectories where $N_{L}$ is of the same order of magnitude as was used for one-particle simulations in \cite{UsMonteCarlo}). The reason for this is the larger volume of phase space available for the particles to explore, making obtaining \textit{representative samples} of the space of trajectories such that the particles resolve \textit{one another} more numerically challenging. Since, however, the generated trajectories are independent and there is no requirement to store the trajectories in memory, this increased computational burden could be distributed naturally on a HPC (though we repeat that the simulations are not memory-intensive). 
    
    As we pointed out when developing the multi-particle WMC formalism, our approach generates trajectories with respect to free-particle distributions on velocities, which as such do not incorporate any information on the systems' potentials. Previous studies have shown that this typically leads to an undersampling problem for late times, which can be overcome by introducing fictitious potentials to constrain the spatial extent of the trajectories and then compensating for that bias \cite{UsPv}. This could be achieved here, in section \ref{sec:HO}, by modifying the algorithms for trajectory generation to incorporate the harmonic interaction, or by following the more familiar approach of simulating trajectories taking the full action into account (e.g.~through heat bath thermalisation or using the Metropolis-Hastings algorithm). This would then allow for a direct estimation of operator expectation values from an ensemble of trajectories distributed according to the Boltzmann weight of the full (i.e. interacting) theory.
    	
    One important and natural avenue for future work would be the incorporation of spin, which can conveniently be described by the use of Grassmann variables on the particle worldlines. Work towards a Monte Carlo simulation of these DoFs in the spirit of WMC is in progress. In taking this step, we would also need to incorporate the fermion statistics for indistinguishable identical particles in the trion systems discussed in section \ref{sec:Trions} -- this would need to be implemented at the level of the path integral trajectories, so needing further modification of the worldline algorithms. The proof of principle demonstrated here also shows that we may return to the case of relativistic quantum mechanics which -- via the worldline formalism -- will allow application of WMC to multi-particle systems in the context of QFT, and perhaps improved studies of bound state formation in a field theoretic framework. 
    
    It would also be interesting to study the performance of the WMC compared to the standard Monte Carlo computation of the path integral~\cite{Creutz:1980gp, Mittal:2018gte} (using Metropolis-Hastings as a means of thermalisation to a distribution in paths constructed using the \textit{full} action) as well as exploring machine learning based methods~\cite{Chen:2022ytr} or quantum algorithms~\cite{Georgescu:2013oza}. This would naturally encourage increasing the number of particles in the systems beyond the two- or three-particle systems studied here, to few bosonic of fermionic particles in a confining trap -- we would expect the WMC method to be perfectly suitable for studying such systems, although the computational processing requirements would make this feasible on a suitably resourced HPC.

    % \J{In the context of applications to condensed matter, this work could be extended by .....\\
    % Please add comments here on what future work could do -- it may be relevant to think about simulating wavepacket evolution or to improve how realistic the interaction and / or external potentials are to better model an exciton system. Or perhaps we look at (non-equal!) effective masses?}
	
	%%%%%%%%%%%%%%%%%%%%%%%%%%%%%%%%%%%%%%%%%%%%%%%%%%%%%%%%%%%%%%%%%%%%%%%%%%%%%%%%%%%%%%%%%%%%%
	\bigskip
	\section*{Acknowledgements}
	The authors thank Michael Faulkner and Rachel Kane for helpful conversations about Monte Carlo sampling (particularly deterministic processes) in statistical physics.\\
	Part of this work has been carried out while JPE was stationed at the Università degli studi di Modena
	e Reggio Emilia, under the “Visiting professor 2023-2024” scheme.
	This work was carried out using the computational facilities of the High Performance Computing Centre, University of Plymouth -- www.plymouth.ac.uk/about-us/university-structure/faculties/science-engineering/hpc. IA is funded by a University of Plymouth doctoral scholarship.
	
	%%%%%%%%%%%%%%%%%%%%%%%%%%%%%%%%%%%%%%%%%%%%%%%%%%%%%%%%%%%%%%%%%%%%%%%%%%%%%%%%%%%%%%%%%%%%%
	%\newpage
	\bigskip
	\bigskip
	\appendix
	\section{Monte Carlo algorithms}
	\label{sec:ApAlg}

		In our simulations we have adopted the ``Yloop'' algorithm presented in \cite{UsMonteCarlo}, where a direct diagonalisation of the argument of the exponent in \eqref{eq:GaussianDistMulti} is achieved on the level of the variables $q_{k}$. According to its developers, the Yloop algorithm has fewer algebraic steps than an earlier ``vloop'' algorithm (presented in \cite{Gies:2003cv}) and is nearly 10\% more efficient than the latter. We note in passing that the vloop algorithm was created to work in the QFT context, where the loops are required to have a centre of mass fixed at zero -- the YLoop algorithm was also originally used in this setting but was later employed in simulations adapted to quantum mechanics \cite{UsMonteCarlo}.
		
		For quick reference, we show here the Yloop algorithm generating unit loops for a single particle, following \cite{UsMonteCarlo}. Denoting by $Y$ the sum in the exponent of \eqref{eq:GaussianDistMulti}, i.e., $Y = \sum_{k=1}^{N_p}(q_{k}-q_{k-1})^2$, it can be non-orthogonally diagonalised by
		\begin{equation}
			Y = \sum_{k=1}^{N_{p}-1}\frac{N_{p}+1-k}{N_{p}-k}\,\bar{q}_{k}^{2},
		\end{equation}
		where
		\begin{equation}
			\bar{q}_{k} = q_{k} - \frac{N_{p}-k}{N_{p}+1-k}\,q_{k-1},\quad k=1,2,\ldots, N_{p}-1.
		\end{equation}
		Then, the corresponding algorithm reads as follows:
		\begin{enumerate}
			\item Generate $N_{p}-1$ vectors $\omega_{i}$, $i=1,2,\ldots,N_{p}-1$, distributed according to $\mathscr{P}(\omega_{i}) \propto \e^{-\omega_{i}^2}\,.$
			\item Compute
			\begin{equation}
				\bar{q}_{i} = \sqrt{\frac{2}{N_p}}\sqrt{\frac{N_{p}-i}{N_{p}+1-i}}\,\omega_{i},\;i=1,2,\ldots,N_{p}-1.
			\end{equation}
			\item Construct the unit loop according to 
			\begin{align}
				\nonumber q_1 &= \bar{q}_1\,, \\
				q_i &= \bar{q}_i + \frac{N_{p}-i}{N_{p}+1-i}\,q_{i-1},\; i=2,3,\ldots,N_{p}-1.
			\end{align}
			\item Repeat the process $N_{L}$ times.
		\end{enumerate}
The configuration space worldline trajectory is then obtained by rescaling the unit loop according to (\ref{eq:xUnit}).		
		
		It is straightforward to adapt the Yloop algorithm for multi-particle systems since it produces worldlines according to the free particle weight on velocities, (\ref{eq:GaussianDistMulti}). This factories into an independent kinetic part associated to each particle,
		\begin{equation}
			Y \rightarrow \sum_{j=1}^{n}Y_{j}
		\end{equation}
		where $Y_{j}$ is the free particle exponent for particle $j$. As such we simply apply the algorithm once for each particle in the ensemble, adding a step:
		\begin{enumerate}\setcounter{enumi}{4}
			\item Repeat steps 1.--4. $n$ times, to produce $q_{j,i}$ for $j = 1, 2, \ldots n$.
		\end{enumerate}
Note, however, that there is no fundamental obstacle to including terms that mix the particle number as long as these remain quadratic in the $q_{j,i}$. In such cases the multi-particle expression for $Y$ can still be diagonalised with respect to these variables, trajectories generated in this diagonal basis and then be separated into real-space worldlines. This could be useful, for example, if we wished to introduce harmonic inter-particle interactions along the lines of \cite{UsPv} so as to avoid the late time undersampling problem (we discuss this in the conclusion section). Moreover, since the algorithm produces mutually independent and identically distributed trajectories, it can be distributed across multiple cores -- for this we used the High Performance Computing (HPC) cluster ``Lovelace'' at the University of Plymouth.

This is clearly a simple enough modification to the WMC algorithms. Hence this manuscript does not focus on the algorithms or their implementation but rather on benchmarking their application to condensed matter systems -- both with respect to their predictions of the ground state energies of various systems and, for the first time, giving a systematic analysis of their scaling behaviour and computation time.

Simplified versions of the code used for two and three particles are available at \href{https://github.com/IAhumada8/Multi-particle-WMC.git}{github.com/IAhumada8/Multi-particle-WMC.git}

	%%%%%%%%%%%%%%%%%%%%%%%%%%%%%%%%%%%%%%%%%%%%%%%%%%%%%%%%%%%%%%%%%%%%%%%%%%%%%%%%%
	\section{Bare Coulomb potential and the smoothing process for two particles}
	\label{sec:ApSmooth}
	
		In this Appendix, we present a generalisation of the smoothing process described in \cite{UsMonteCarlo} for an interaction between two particles with a short-distance singularity. In particular, when $d=0$ the soft-Coulomb potential reduces to the well known \textit{bare} Coulomb potential and exact solutions exist in 1D \cite{RLoudon1959Hydrogen1D,RLoudon2016Hydrogen1D}, 2D \cite{Yang:1991} and 3D \cite{Griffiths:2018}. We have that the energies for two particles, in low dimensions and in our units, are 
		\begin{align}
			\nonumber E_i &= -\frac{\mu \alpha^{2}}{2i^{2}},\quad\text{for 1D}\\
			E_l &= -\frac{\mu \alpha^{2}}{2(l-\frac{1}{2})^{2}},\quad\text{for 2D}
		\end{align}
		with $i=0,1,2,\ldots$, $l=1,2,\ldots$. So, for the ground state, we have that the energy is ill defined in 1D \cite{Yepez:2011,Gebremedhin:2014,Carrillo:2015} ($E_0=-\infty$) and in 2D is $E_0=-2\mu\alpha^{2}$.
		
However, if we want to simulate the limit when $d=0$ using \eqref{eq:Soft-Coulomb} we tend to encounter large deviations from the expected (late-time) monotonic behaviour for the logarithm of the propagator. This issue was addressed in \cite{UsMonteCarlo} for singular potentials of one particle, where these large deviations were named \textit{skyscrapers}. These skyscrapers are caused by some paths in the ensemble passing close to the singularity, thus giving spurious contribution to the propagator. To correct these discrepancies, a \textit{smoothing} procedure was proposed in the same article for one particle that approaches the origin in the presence of a static Coulomb potential. Here we present the procedure extended to two particles, where the singularity exists when their relative separation becomes small. 
		
The main idea of the smoothing process is to evaluate analytically the line integral of the Wilson line for our discretised trajectories. To do this we must assume some functional form for the trajectory propagation between the discretised points $\{x_{i,j}(\tau_{k})\}$. For simplicity, this has previously been done assuming the particles propagate along straight lines between these points, which we will continue to employ here\footnote{This is clearly a drastic approximation, since Brownian motion trajectories should be almost nowhere differentiable and linear interpolation between points on a ``true'' trajectory given a-priori would obviously miss curvature corrections in the line integral determining the Wilson line. However, the Gaussian distribution on velocities seems to allow this severe approximation, with errors in $1/N_{P}$ being negligible compared to the statistical error in approximating the path integral with a finite number of loops for precisely the trajectories that contribute significantly to the path integral.}.
		
To show this, we consider the Wilson line for the bare Coulomb potential, $W(v)=\e^{-v}$, with $v$ defined as
		\begin{align}
			\nonumber v &\equiv -T\int_{0}^{1}du\,\frac{\alpha}{r}, \quad \mbox{where} \\
			\nonumber r &= |x_1(u)-x_2(u)| \\
			\nonumber &= \Big|\Big(x_1+(x_1'-x_1)u+\sqrt{\frac{T}{m_1}}q_1(u)\Big) \\
			&- \Big(x_2+(x_2'-x_2)u+\sqrt{\frac{T}{m_2}}q_2(u)\Big)\Big|\,.
		\end{align}
		Here, $x_i$ and $x_i'$, $i=1,2$ are the initial and final space points respectively.
		Its discretised version is
		\begin{align}
				\nonumber v &= -\frac{T}{N_p}\sum_{k=1}^{N_p}\frac{\alpha}{r_k} \,, \\ 
			% \nonumber r_k &= \Big|\Big(x_1+(x_1'-x_1)\frac{k}{N_p}+\sqrt{\frac{T}{m_1}}q_{1,k}\Big) \\
			% &- \Big(x_2+(x_2'-x_2)\frac{k}{N_p}+\sqrt{\frac{T}{m_2}}q_{2,k}\Big)\Big|\,,
			\nonumber r_k &= \big| x_{1}(\tau_{k}) - x_{2}(\tau_{k}) \big|\,,
		\end{align}
		where $r_k \equiv r(\tau_{k})$ is evaluated at the k$^{\rm th}$ point in the discretised path. 
		In analogy with the smoothing process for one particle, we parameterise straight line paths between consecutive points along the trajectories by (a shared parameter due to maintain locality) $l\in[0,1]$, with respect to which the lines joining $x_{1,k-1}$ to $x_{1,k}$ and $x_{2,k-1}$ to $x_{2,k}$ can be written as
		\begin{align}
			\nonumber x_{1,k}(l) &= x_{1,k-1}+(x_{1,k}-x_{1,k-1})l \,, \\
			\nonumber x_{2,k}(l) &= x_{2,k-1}+(x_{2,k}-x_{2,k-1})l \,;.
		\end{align}
		With this, the inter-particle separation at any point along these lines is
		\begin{widetext}
			\begin{align}
				\nonumber &|x_{1,k}(l)-x_{2,k}(l)| =\\
			% \sqrt{(x_{1,k}(l)-x_{2,k}(l))\cdot (x_{1,k}(l)-x_{2,k}(l))}\\
				 &\sqrt{(x_{1,k-1}-x_{2,k-1})^2+2\{(x_{1,k-1}-x_{2,k-1})[(x_{1,k}-x_{2,k})-(x_{1,k-1}-x_{2,k-1})]\}l + [(x_{1,k}-x_{2,k})-(x_{1,k-1}-x_{2,k-1})]^2l^2} \,.
			\end{align}
			
			With this parametrisation we can compute $\int_{0}^{1}\frac{dl}{r}$ analytically as
			
			\begin{align}
				\nonumber \int_{0}^{1}&\frac{dl}{|x_{1,k}(l)-x_{2,k}(l)|} = \frac{1}{|(x_{1,k}-x_{2,k})-(x_{1,k-1}-x_{2,k-1})|}\\
				&\quad \times \ln\Big|\frac{(x_{1,k}-x_{2,k})^2-(x_{1,k}-x_{2,k})\cdot(x_{1,k-1}-x_{2,k-1})+|(x_{1,k}-x_{2,k})-(x_{1,k-1}-x_{2,k-1})||x_{1,k}-x_{2,k}|}{-(x_{1,k-1}-x_{2,k-1})^2+(x_{1,k}-x_{2,k})\cdot(x_{1,k-1}-x_{2,k-1})+|(x_{1,k}-x_{2,k})-(x_{1,k-1}-x_{2,k-1})||x_{1,k-1}-x_{2,k-1}|}\Big| \,.
				\label{eq:smoothed}
			\end{align}
		\end{widetext}
		
\begin{figure}
			\centering
			\includegraphics[width=0.5\textwidth]{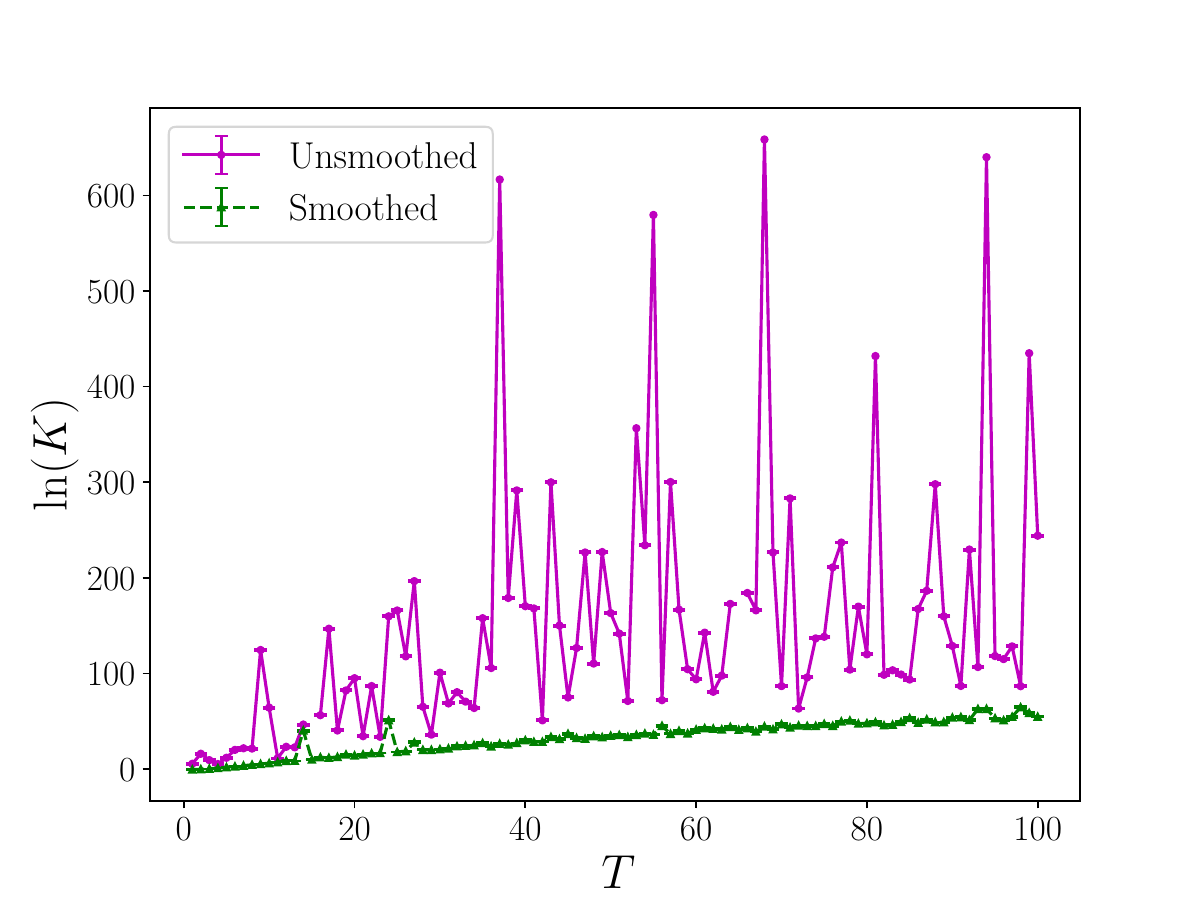}
			\caption{Comparison of $\ln(K)$ before and after smoothing for two particles interacting via the Coulomb potential in 2D. We used $N_L=2000^2$ loops, $N_p=5000$ points per loop, $m_{1}=m_{2}=1$ and $\alpha=1$. Discontinuities represent Nan values in the unsmoothed estimate.}
			\label{Fig:Smoothing}
		\end{figure}

		In Fig.~\ref{Fig:Smoothing} we show the logarithm of the propagator for the Coulomb potential and can observe the large deviations (skyscrapers) present in the estimate of this kernel. We also show the same estimate after the application of the smoothing process. We can see that the skyscrapers have been removed from the estimation, although some smaller scale deviations from the desired linear behaviour remain. This is to be expected, because although the smoothing procedure has removed the artificially strong linear ($1/r$) divergence, (\ref{eq:smoothed}) is left with a (true) logarithmic divergence whenever the quantity $(x_{1,k}-x_{2,k})$ is close to zero or these points are such that the straight line joining them passes through the origin.
		
		%Applying the above smoothing process we can estimate the ground state energy for two particles interacting via the bare Coulomb potential in 2D. 
		However, smoothing the divergences allows us to find a linear region and estimate the ground state energy for two particles interacting via the bare Coulomb potential in 2D. Using $m_{1}=m_{2}=\alpha=1$, $N_{L}=5000^{2}$ and $N_{p}=10000$ we estimate the ground state energy to be $E_0=-0.9868(6) \pm 0.02(2)$, which is in close agreement  (percent error of $\approx 1.32\%$) with the analytical result $E_0=-1$. This WMC estimation gives a better result that the one provided by a one-particle diagonalisation code: $E_0=-0.9805(4)$ (percent error of $\approx 1.946\%$) with a grid of size $N_{T}=1000^{2}$ total points.

	\section{Time complexity and convergence analysis}
	\label{sec:ApTimeScaling}
    	In this appendix we provide results obtained from a systematic exploration of how the simulation times of the WMC and diagonalisation code scales with the appropriate discretisation parameters and number of spatial dimensions. We also study the convergence of the diagonalisation code for systems in two dimensions and we explicitly show the Pad\'e approximants used to estimate the errors of this method. 
		
		Numerical methods based on the direct resolution of the discrete version of the time-(in)dependent Schr\"odinger equation are characterised by a significant increase in computation time and memory storage requirements as the dimensionality of the system to be studied increases. Here we illustrate how the diagonalisation method scales in time compared to the WMC formalism when used to estimate the ground state energy of two particles interacting via the soft Coulomb potential and each particle subjected to an external Gaussian potential, the reference system studied in section \ref{sec:softCoulombGaussianPotentials}. 
		
		For the diagonalisation method, we measure the running time of a  total number of grid points $N_{T}=N^{2D}$ ($N$ grid points for each particle in each dimension), ranging from $N_{T}=10^{2D}$ to $N_{T}=100^{2D}$, in increments of $10^{2D}$ in 1D and 2D; in the 3D case  it was not possible to use a number of total grid points $N_{T}>22^{6}$ due to the physical memory limitations of our hardware; therefore we limit ourselves to performing simulations for $N\in[10,22]$ with increments of 2. For the WMC method, we fix the number of points per loop $N_{p}=5000$ and calculate the execution time for a number of loops ranging from $\tilde{N}_{L}=10^{5}$ to $\tilde{N}_{L}=10^{6}$ for every 100,000 loops. Tests were run using all the resources of a LENOVO MT 30FN BU Think FM ThinkStation P360 Tower with a 12th Gen Intel(R) Core(TM) i7-12700, running at 2100 MHz with 20 logical processors; the Tower has 32 GB of physical memory (RAM).		
		
\begin{figure}
			\centering
			\includegraphics[width=0.5\textwidth]{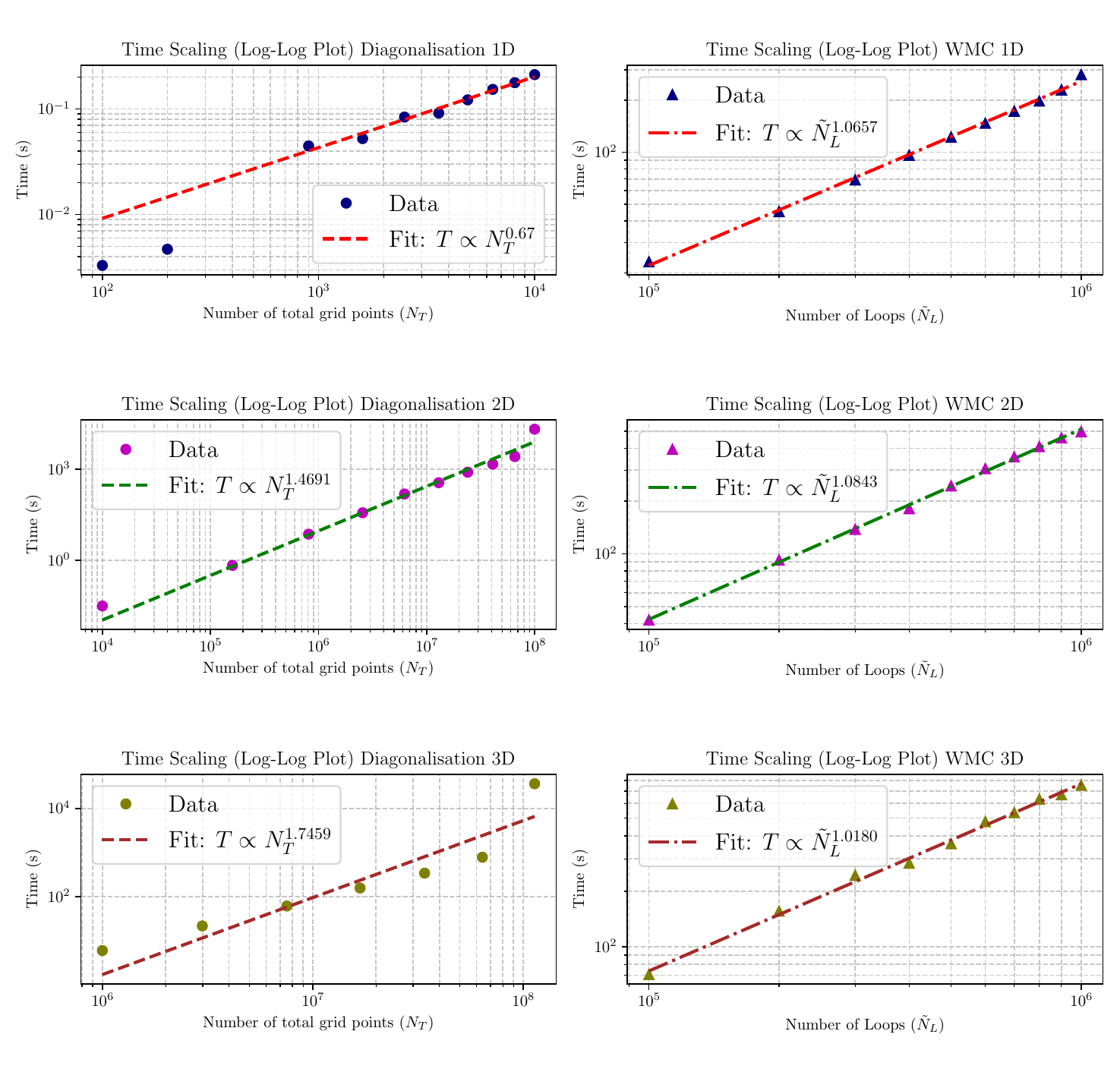}
			\caption{Time execution as a function of the number of total grid points $N_{T}$ or number of loops $\tilde{N}_{L}$ (diagonalisation code or WMC respectively) in 1D, 2D and 3D in a log-log space.}
			\label{Fig:Time_scaling-1}
		\end{figure}

		Figure \ref{Fig:Time_scaling-1} shows the runtime results in seconds for both methods as a function of the number of total grid points or the number of loops on dual logarithmic axes and sequentially in one, two and three dimensions. There we fit a first degree polynomial $\log(T) = a + b\log(N_{(T,L)})$ to capture the large $N_{(T,L)}$ behaviour (discarding points for smaller values of $N_{T}$ not compatible with a linear fit). This fit for simulation time as a function of discretisation parameter is also shown in the plot. From this fit, the simulation time of the WMC formalism is consistently close to linear with $\tilde{N}_{L}$ regardless of the number of spatial dimensions, well approximated by $\mathcal{O}(\tilde{N}_{L})$ for fixed $N_{p}$. This is because the trajectory generation is sequential and memoryless. The fits show the diagonalisation method to be a polynomial time algorithm where the time complexity varies depending on the dimensionality of the system, being approximately $\mathcal{O}(N_{T}^{0.67})$, $\mathcal{O}(N_{T}^{1.47})$ and $\mathcal{O}(N_{T}^{1.75})$ for one, two and three dimensions respectively.
		
\begin{figure}
			\centering			\includegraphics[height=0.25\textheight, trim={2cm 0 0 0}]{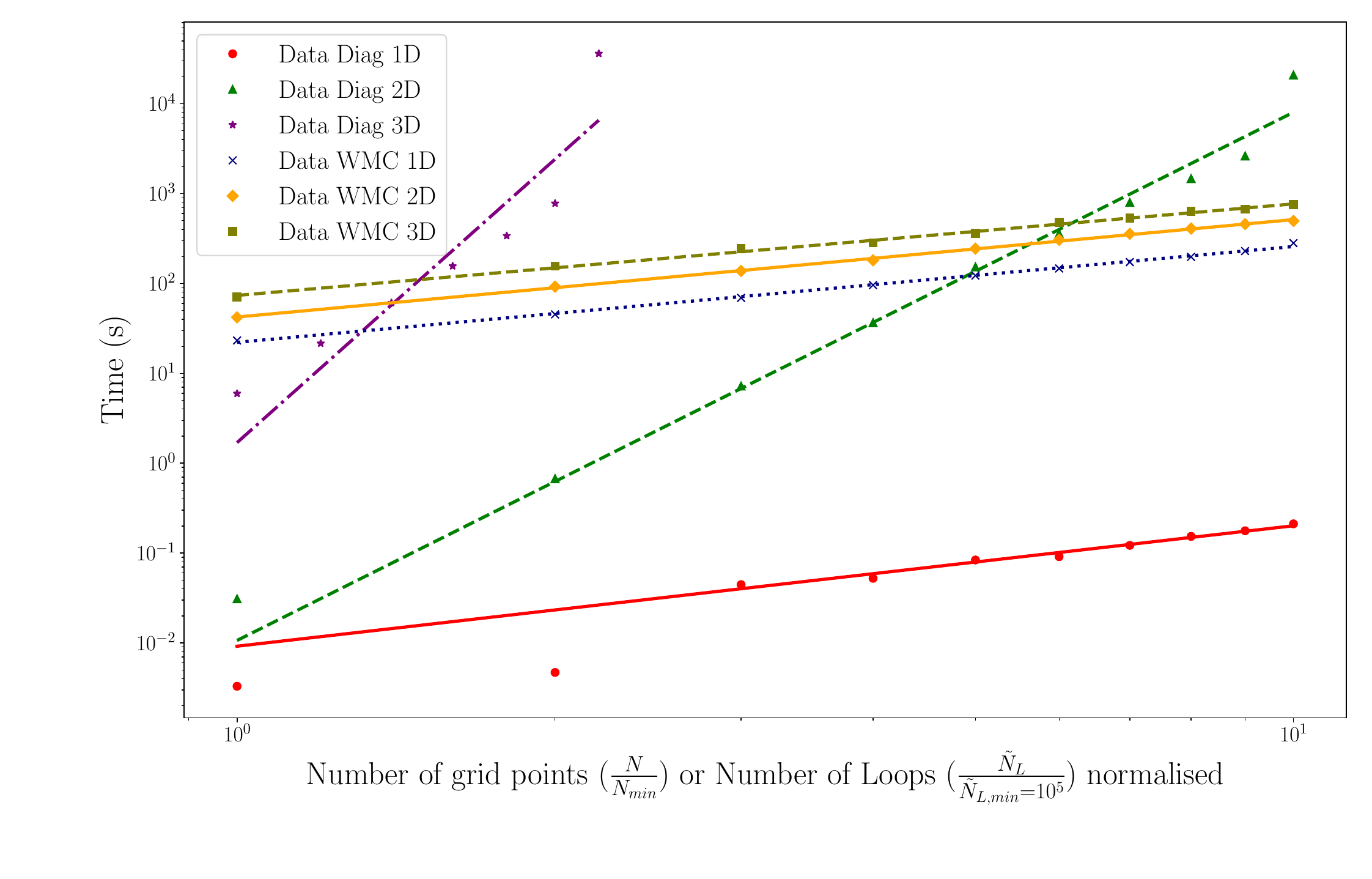}
			\caption{Comparison in execution time as a function of the number of grid points $N$ or number of loops $\tilde{N}_{L}$ normalised with their minimum value in a log-log space. We refer to Fig.~\ref{Fig:Time_scaling-1} to see the values of the fitted lines.}
			\label{Fig:Time_scaling-2}
		\end{figure}

		In Fig.~\ref{Fig:Time_scaling-2} we make a direct comparison of the two methods by normalising the number of grid points (per particle per dimension) and number of loops against their smallest values, i.e., $N_{\text{min}}=10$ and $\tilde{N}_{L,\text{min}}=10^5$ respectively. This makes clear that the diagonalisation method is faster (in absolute terms) than the WMC formalism in 1D for sparser discretisation / number of loops. This feature has consistently allowed us to increase the number of grid points beyond 100 without compromising the computation time and memory requirements. However, this is not the case in two dimensions where the execution times are comparable with those of the WMC formalism already for $N_{T}=60^{4}$, and for finer grids the method scales less favourably than WMC. It is at this point that the strong time and memory requirements in the diagonalisation method manifest themselves, for example, for $N_{T}=100^{4}$ the computation requires about $2\times10^{4}$ seconds to complete and about 25 GB of physical memory, compared to 500 seconds and 400 MB (20 MB per core) for $\tilde{N}_L=10^{6}$ in the WMC formalism. 		Finally, for the 3D case the memory requirements and computational complexity make it prohibitively expensive to use $\gtrsim N = 20$ discretisation points per dimension in our Workstation. 
        
        The rising cost of the calculation with dimension is an example of the curse of dimensions (see for example \cite{Hartung:2020nci}) The increasing cost of solving quantum mechanics problems as the number of particles increased was the motivation for Feynman to develop quantum computers \cite{Feynman:1981tf}.
		
		%\vspace{6mm}

	   \subsection{Convergence of ground state energy estimations for the diagonalisation method}
	
		In this subsection we analyse the convergence of the diagonalisation method in calculating the ground state energy as a function of the number of grid points per particle per dimension $N$ for the systems studied in section \ref{sec:softCoulombGaussianPotentials} and \ref{sec:softCoulombSqrWells} in two dimensions. In addition, we explicitly show the Pad\'e approximants for each case and the estimated asymptotic value of $E_{0,\infty}$ when we extrapolate $N \to \infty$.

		\begin{figure}[b]
			\centering
	        \includegraphics[height=0.295\textheight]{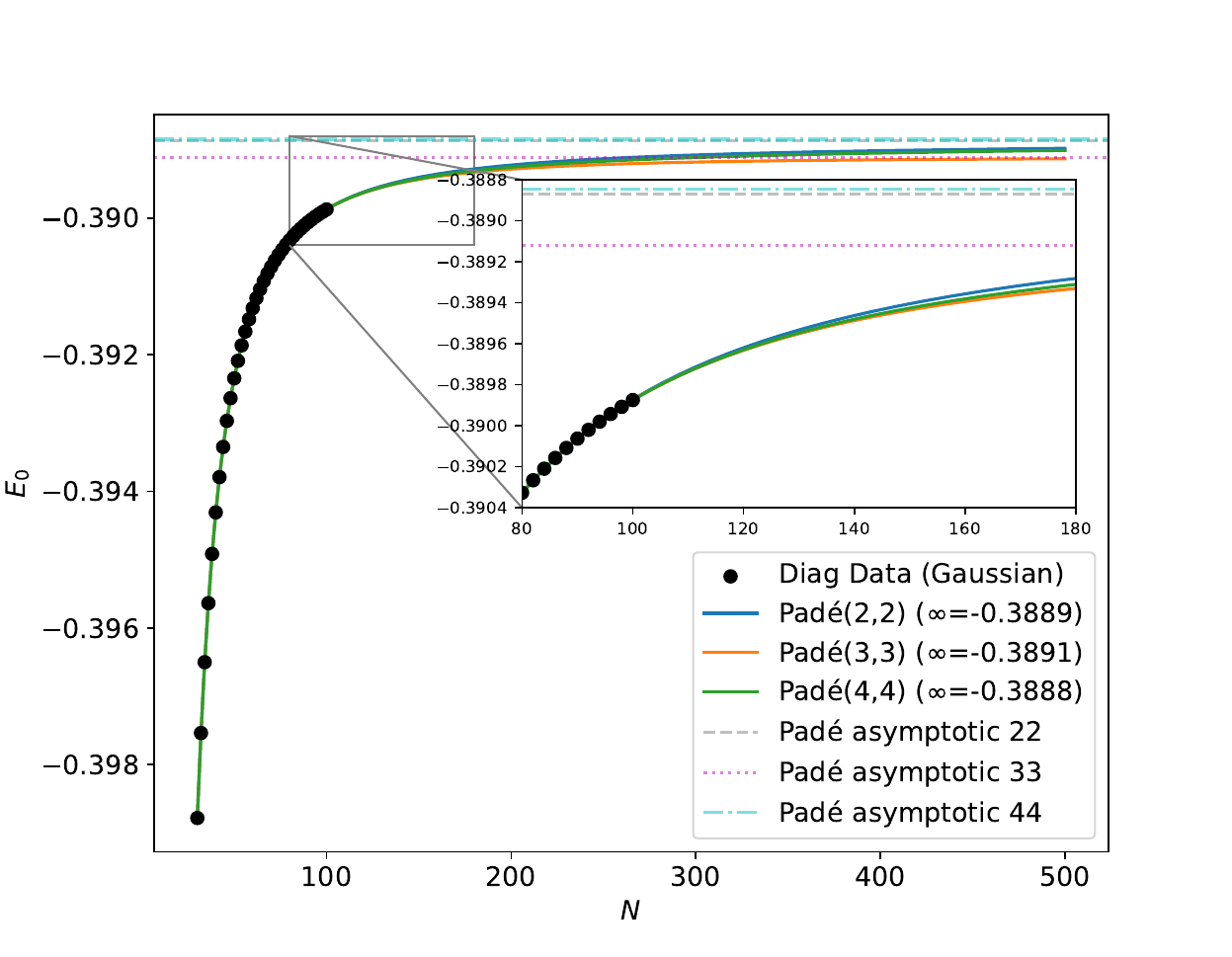}
			\caption{Convergence plot for the soft-Coulomb potential + Gaussian potentials with three different Pad\'e approximants $(m,n)$. The horizontal lines denotes the estimated asymptotic values of the ground state energy when we take the limit $N \to \infty$. Inset: zoom to the region for values of $N\in[80,180]$.}
			\label{Fig:Pade_GaussPot}
		\end{figure}
		
		We start by analysing the convergence of an interacting electron-hole pair through the soft-Coulomb potential for a fixed value of distance $d=1$, where the electron(hole) is subject to a Gaussian potential well(barrier). We perform simulations for a number of grid points $N$ over a range of $N \in [30,100]$ with steps of $2$ and fit three different Pad\'e approximants corresponding to the cases $m=n=2,3,4$. From Fig.~\ref{Fig:Pade_GaussPot} we can observe how the results converge in a monotonically increasing manner, where the points get closer to each other as we increase the number of grid points used $N$. The Pad\'e approximants fit the data points well and only a slight deviation between them is seen when we start to approach the asymptotic limit, marked as the horizontal lines for each case. We notice that with the diagonalisation method we \textit{underestimate} the ``real value'' for the ground state energy of this system, a fact that is verified by the WMC formalism where each point of this method in Fig.~\ref{Fig:2DIXNonSeparableSys} is above the results generated by the diagonalisation one. To estimate the error associated with the diagonalisation method, we use for this system the Pad\'e approximant corresponding to $(m=4,n=4)$ because we noticed an agreement with the case $(2,2)$ in the third significant figure for the asymptotic value of $E_{0,\infty}$.

\begin{figure}
			\centering
			\includegraphics[height=0.25\textheight, trim={0 0 0 1.5cm}]{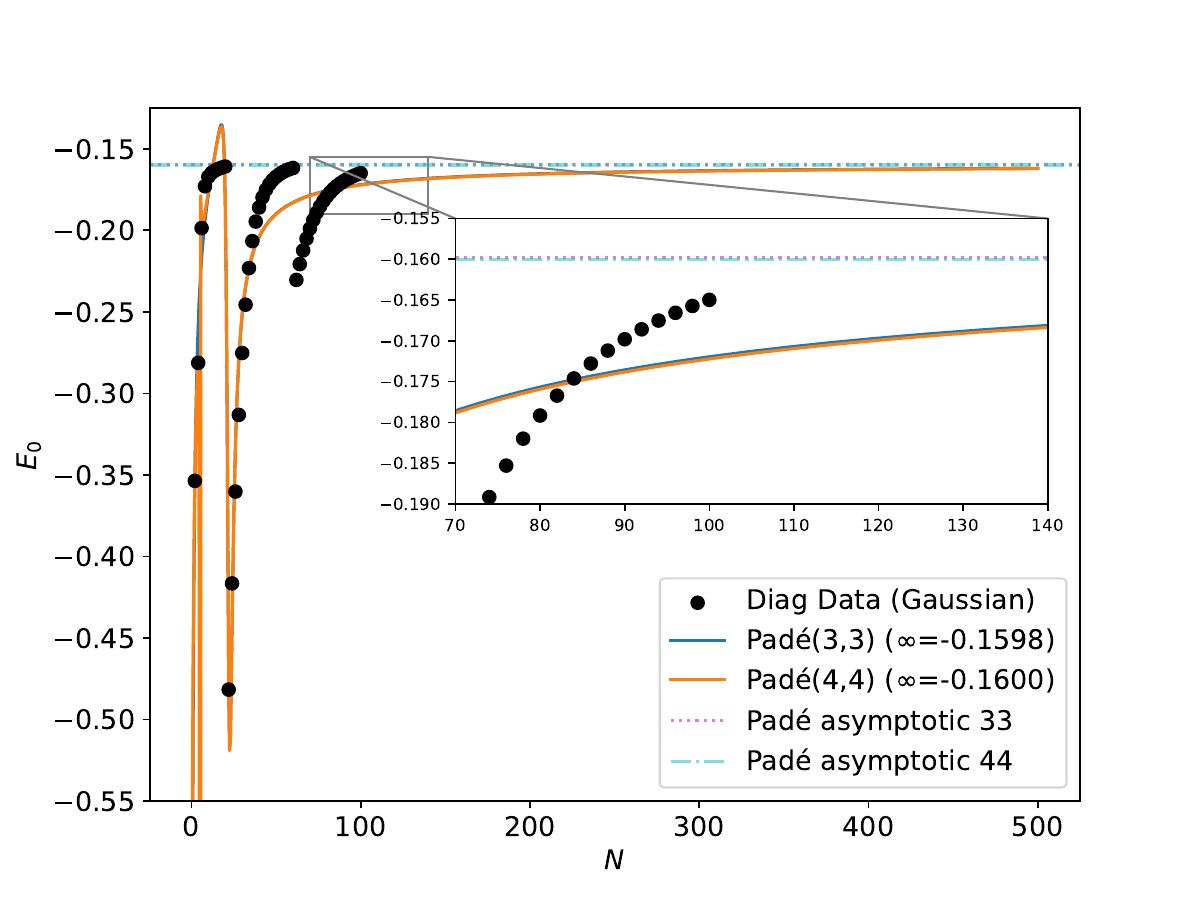}
			\caption{Convergence plot for the soft-Coulomb potential + Quantum wells with two different Pad\'e approximants $(m,n)$. The horizontal lines denotes the estimated asymptotic values of the ground state energy when we take the limit $N \to \infty$. Inset: zoom to the region for values of $N\in[70,140]$.}
			\label{Fig:Pade_SqrWells}
		\end{figure}

		The second system we study is the case of an electron-hole pair interacting through the soft-Coulomb potential for a fixed value of the distance $d=2.75$, where the electron(hole) is subject to a finite quantum potential well (barrier) of widths $L_{x}=L_{y}=1$ and depth (height) $V_{e}=-1$ ($V_{h}=1$). For this case, we perform simulations for a number of grid points $N$ over a range of $N \in [2,100]$ with steps of $2$ and fit two different Pad\'e approximants corresponding to the cases $m=n=3,4$. In this case, we can see from Fig.~\ref{Fig:Pade_SqrWells} that the behaviour of the data points is no longer monotonically increasing, but rather an increasing trend followed by ``jumps'' is observed as we increase the value of $N$. We associate this behaviour with the difficulty of capturing the sharp edges of the square potentials due to the spatial discretisation of the grid. As we increase the resolution of the grid, i.e., $N$, the discontinuities become increasingly smaller in magnitude because we capture information from the edges with greater precision and the data points are expected to converge to a given value. We estimate this value by fitting Pad\'e approximants again for values $m=n=3,4$; here the curves fit well for the first points and are able to follow the first discontinuity of the data points but are not able to follow the second. However, the approximants capture the trend of the data points as we increase $N$. 
		
		Following studies of the convergence of the previous system, we choose the Pad\'e approximant corresponding to $(3,3)$, which gives an upper bound since we know that the diagonalisation method tends to underestimate the values for the ground state energy. Even so, the difference between both approximants occurs around the fourth significant figure, sufficient precision for our purposes of estimating the order of magnitude of the error bars and creating the shaded region that indicates this discrepancy.

	\bibliographystyle{apsrev4-2}
	\bibliography{bibWLMC.bib}
\end{document}